\documentclass[useAMS,usenatbib]{aa}
\usepackage{graphicx}
\usepackage{natbib}
\usepackage{txfonts}
\usepackage[normalem]{ulem}
\begin{document} 

\title{Observing multiple stellar populations with FORS2@VLT} 
\subtitle{Main sequence photometry in outer regions of NGC\,6752, NGC\,6397, and NGC\,6121 (M~4).\thanks{Based on observations at the European Southern Observatory using the Very Large Telescope on Cerro Paranal through ESO programme 089.D-0978 (P.\,I.\,G.\,Piotto)}}


\author{D.\,Nardiello\inst{1,2,3}
  \and
  A.\,P.\,Milone\inst{2}
  \and
  G.\,Piotto\inst{1,3}
  \and
  A.\,F.\,Marino\inst{2}
  \and
  A.\,Bellini\inst{4}
  \and
  S.\,Cassisi\inst{5}
}
\institute{Dipartimento di Fisica e Astronomia ``Galileo
  Galilei'', Universit\`a di Padova, Vicolo dell'Osservatorio 3, I-35122
  Padova, Italy\\
  \email{domenico.nardiello@studenti.unipd.it; giampaolo.piotto@unipd.it}
  \and
  Research School of Astronomy and Astrophysics, The Australian National University, Cotter Road, Weston, ACT, 2611, Australia \\
  \email{antonino.milone@anu.edu.au; anna.marino@anu.edu.au}
  \and
  INAF -- Osservatorio Astronomico di Padova, Vicolo dell'Osservatorio 5, I-35122, Padova, Italy \\
  \and
  Space Telescope Science Institute, 3700 San Martin Dr., Baltimore, MD\,21218, USA \\
  \email{bellini@stsci.edu}
  \and
  INAF -- Osservatorio Astronomico di Collurania, Via M. Maggini, I-64100, Teramo, Italy \\
  \email{cassisi@oa-teramo.inaf.it}
}
  
  \date{Received  02 May 2014 / Accepted 23 October 2014}
  
  \abstract{}{ We present the photometric analysis of the external
    regions of three Galactic Globular Clusters: NGC\,6121, NGC\,6397
    and NGC\,6752.  The main goal 
    is the characterization of the multiple stellar populations along
    the main sequence (MS) and the study of the radial trend of the
    different populations hosted by the target clusters.}  {The data
    have been
    collected using FORS2 mounted at the ESO/VLT@UT1 telescope in
    $UBVI$ filters.  From these data sets we extracted high-accuracy
    photometry and constructed color-magnitude diagrams.  We exploit
    appropriate combination of colors and magnitudes which are
    powerful tools to identify multiple stellar populations, like $B$
    versus $U-B$ and $V$ versus $c_{\rm U,B,I}=(U-B)-(B-I)$ CMDs.}{We
     confirm previous findings of
    a split MS in NGC\,6752 and NGC\,6121. Apart from the extreme case of
    $\omega$\,Centauri, this is the first detection of multiple MS
    from ground-based photometry. For NGC\,6752 and NGC\,6121 we compare the number
    ratio of the blue MS to the red MS in the cluster outskirts with
    the fraction of first and second generation stars measured in the
    central regions. There is no evidence for significant radial
    trend.\\
    The MS of NGC\,6397 is consistent with a simple stellar
    population. We propose that the lack of multiple sequences is
    due both to observational errors and to the limited sensitivity of
    $U,B,V,I$ photometry to multiple stellar populations in metal-poor
    GCs. \\
    Finally, we compute the helium abundance for the stellar
    populations hosted by NGC\,6121 and NGC\,6752, finding a  mild
    ($\Delta Y \sim 0.02$) difference between stars in the two
    sequences.}{}
   
   \maketitle
   \section{Introduction \label{intro}}
   Over the last years, the discovery that the color-magnitude
   diagrams (CMDs) of many globular clusters (GCs) are made of
   multiple sequences has provided overwhelming proof that these
   stellar systems have experienced a complex star-formation
   history. The evidence that GCs host multiple stellar populations
   has reawakened the interest on these objects both from the
   observational and the theoretical point of view.
   
   Multiple sequences have been observed over all the CMD, from the
   main sequence (MS, e.g.\,\citealt{2007ApJ...661L..53P}) through the
   sub-giant branch (SGB, e.g.\,\citealt{2012ApJ...760...39P}) and
   from the SGB to the red-giant branch (RGB,
   e.g.\,\citealt{2008A&A...490..625M}) and even in the
     white-dwarf cooling sequence \citep{2013ApJ...769L..32B}.

   Multiple populations along the RGB have been widely studied in a
   large number of GCs (e.g.\,\citealt{2008ApJ...684.1159Y},
   \citealt{2009Natur.462..480L}, \citealt{2013MNRAS.431.2126M}) by
   using photometry from both ground-based facilities and from the
   {\it Hubble Space Telescope} ({\it HST}). In contrast, with the
   remarkable exception of $\omega$\,Centauri
   (\citealt{2007ApJ...654..915S}, \citealt{2009A&A...507.1393B}), the
   investigation of multiple MSs has been carried out with {\it HST}
   only (e.g.\,\citealt{2004ApJ...605L.125B},
   \citealt{2007ApJ...661L..53P}, \citealt{2012ApJ...745...27M} and
   references therein, \citealt{2013ApJ...765...32B}).

   In this paper we will exploit the FOcal Reducer and low dispersion
   Spectrograph 2 (FORS2), mounted at the Very Large Telescope (VLT)
   of the European Southern Observatory (ESO) to obtain accurate {\it
     U},{\it B},{\it V},{\it I} photometry of MS stars in the
   outskirts of three nearby GCs, namely NGC\,6121 (M\,4), NGC\,6397,
   and NGC\,6752, and study their stellar populations.

   The paper is organized as follows: in Sect.~\ref{GCs} we provide an
   overview on the three GCs studied in this paper. The observations
   and the data reduction are described in Sect.~\ref{obs}.  The CMDs
   are analyzed in Sect.~\ref{CMDs}, where we also show evidence of
   bimodal MSs for stars in the outskirts of NGC\,6121 and
   NGC\,6752 and calculate the fraction of stars in each MS. In
   Sect.~\ref{rad} we study the radial distribution of stellar
   populations in NGC\,6752 and NGC\,6121. In Sect. \ref{Helium} we
   estimate the helium difference between the two stellar populations
   of NGC\,6121 and NGC\,6752. A summary follows in Sect.~\ref{sum}.
   \begin{figure*}
     \centering
     \includegraphics[ width=0.32\hsize]{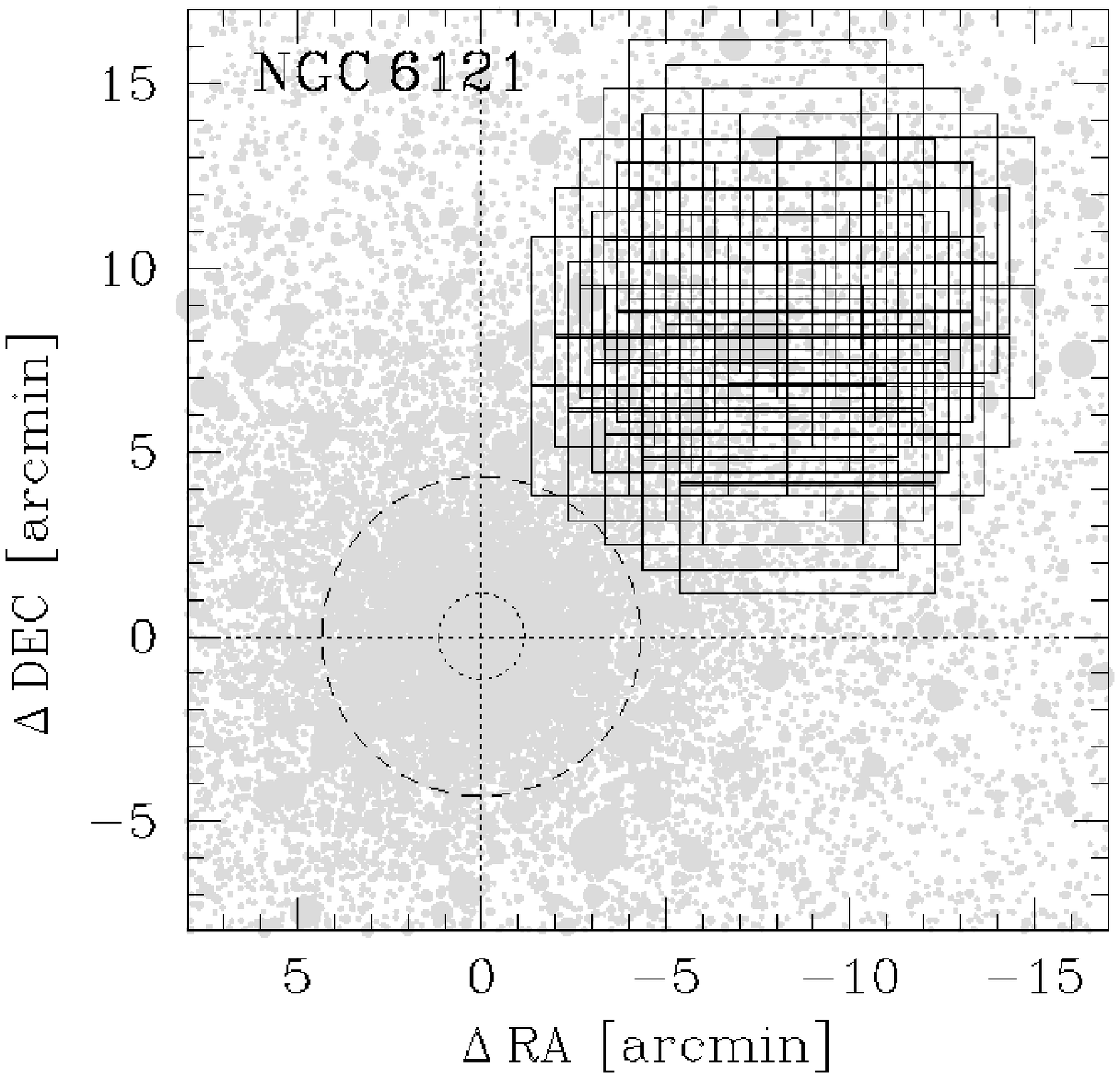} 
     \includegraphics[ width=0.32\hsize]{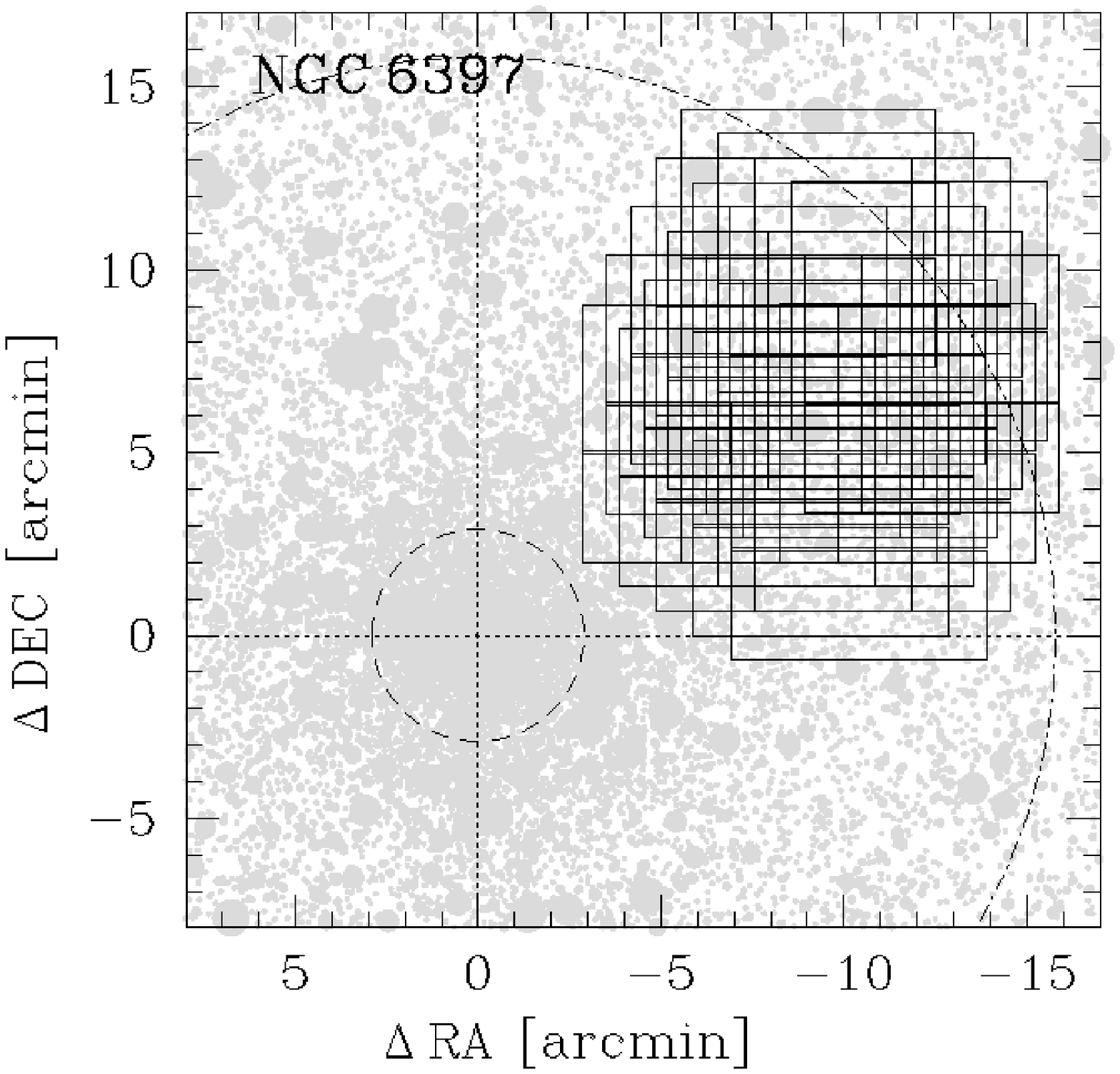} 
     \includegraphics[ width=0.32\hsize]{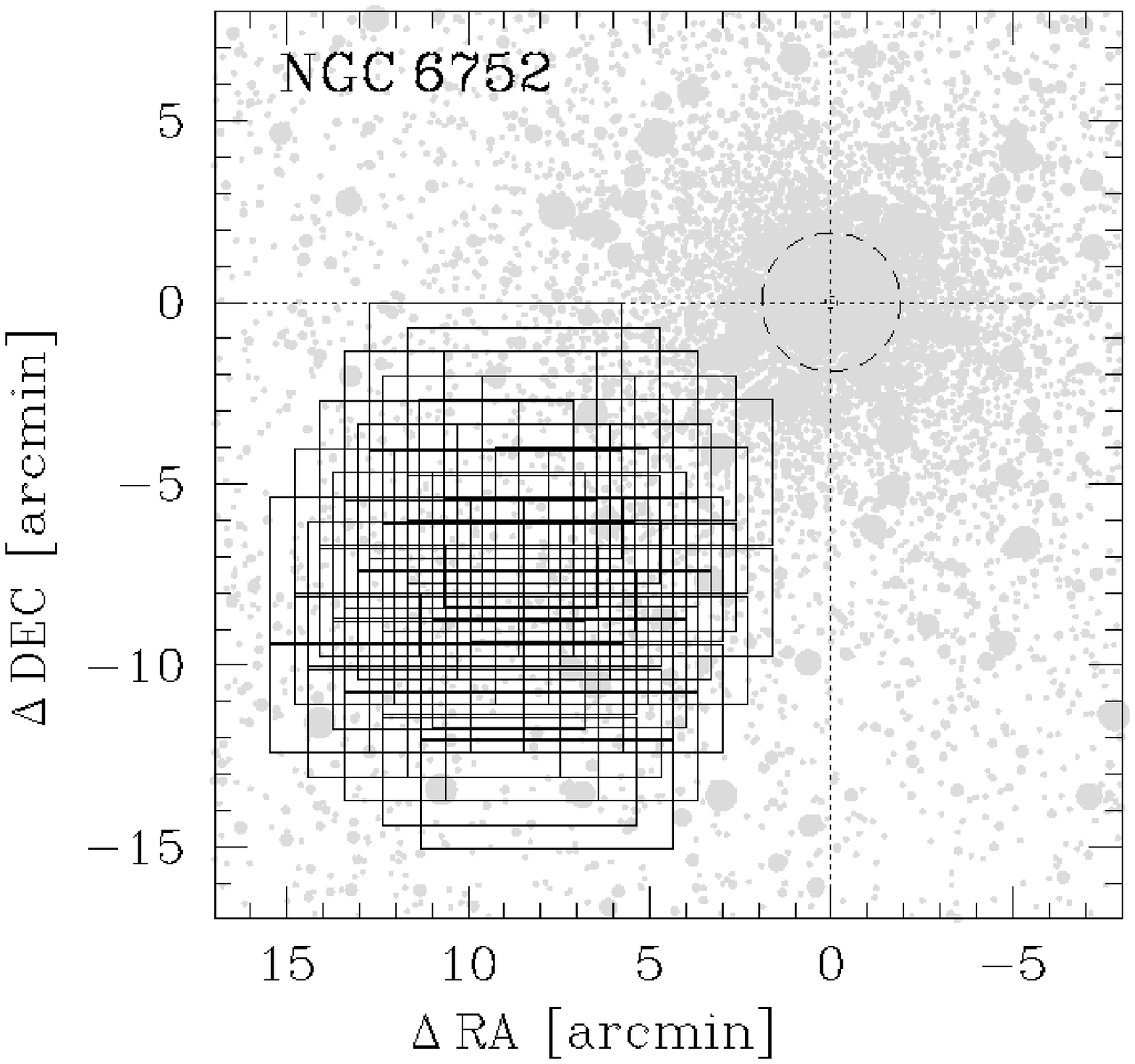} 
     \caption{Dither pattern of the FORS2 images taken in the fields of
       NGC\,6121 (left), NGC\,6397 (middle), and NGC\,6752 (right). The
       dotted, dashed, and dashed-dotted circles indicate the core, the
       half-light, and the tidal radius of the three GCs. NGC\,6397 and
       NGC\,6752 are both post core collapse clusters.}
     \label{footprint}
   \end{figure*}

   \section{Properties of the target GCs\label{GCs}}
   Multiple stellar populations have been widely studied in the three GCs
   analyzed in this paper. In this section we summarize the observational
   scenario and provide useful information to interpret our observations.

   \subsection{NGC\,6121}

   NGC\,6121 is the closest GC ($d\sim2.2$\,kpc) and has intermediate
   metal abundance ([Fe/H]=$-$1.16 \citealt{1996AJ....112.1487H}, 2010
   edition).

   The RGB stars of this cluster exhibit a large spread in the
   abundance distribution of some light-elements such as C, N, O, Na
   and Al (\citealt{1986A&A...169..208G};
   \citealt{1990AJ....100.1561B}; \citealt{1992ApJ...395L..95D};
   \citealt{2005PASP..117..895S}). There is evidence of a CN
   bimodality distribution and Na-O anticorrelation (e.g.\,
   \citealt{1981ApJ...248..177N}; \citealt{1999AJ....118.1273I}).

   The distribution of sodium and oxygen is also bimodal. Sodium-rich
   (oxygen-poor) stars define a red sequence along the RGB in the $U$
   versus $U-B$ CMD, while Na-poor stars populates a bluer RGB sequence
   (\citealt{2008A&A...490..625M}).  Further evidence of multiple
   sequences along the RGB of NGC\,6121 are provided by
   \citet{2009Natur.462..480L} and \citet{2013MNRAS.431.2126M}. NGC\,6121
   has a bimodal HB, populated both on the blue and red side of the
   instability strip. The HB morphology of this cluster is closely
   connected with multiple stellar populations, indeed blue HB stars are
   all Na-rich and O-poor (hence belong to the second stellar generation)
   while red-HB stars have the same chemical composition as
   first-generation ones \citep{2011ApJ...730L..16M}.

   \subsection{NGC\,6397}

   Located at a distance of $2.3$\,kpc, NGC\,6397 is a very metal-poor GC
   ($\mathrm{[Fe/H]}=-2.02$; \citealt{1996AJ....112.1487H}, 2010 edition).

   In late 1970s, \cite{1979ApJ...229..604B} have already demonstrated
   that the RGB stars of this cluster show a spread in light-element
   abundance. NGC\,6397 exhibits modest star-to-star variations of
   oxygen and sodium and a mild Na-O anti-correlation
   (e.g.\,\citealt{2002AJ....123.3277R},
   \citealt{2009A&A...505..117C}). Similarly to NGC\,6121 the
   distribution of sodium and oxygen is bimodal, and the groups of
   Na-rich (O-poor) and Na-poor (O-rich) stars populate two distinct
   RGBs in the Str\"omgren $y$ versus $c_y$ index diagram
   (\citealt{2011A&A...527A.148L}).

   The MS of NGC\,6397 is also bimodal, but the small color separation
   between the two MSs can be detected only when appropriate filters
   (like the F225W, F336W from {\it HST}/WFC3) are used. Observations of
   the double MSs from multi-wavelength {\it HST} photometry have been
   interpreted with two stellar populations with different light-element
   abundance and a modest helium variation of $\Delta Y \sim$0.01 (
   \citealt{2010A&A...511A..70D}, \citealt{2012ApJ...745...27M}).

   \subsection{NGC\,6752} 
   NGC\,6752, is a nearby metal-poor GC ($d=4.0$\,kpc,
   $\mathrm{[Fe/H]}=-1.54$; \citealt{1996AJ....112.1487H}, 2010
   edition).

   Since the 1980s, spectroscopic data reported `anomalies' in the
   light-element abundances of the RGB stars of this cluster
   (\citealt{1981ApJ...244..205N}, \citealt{1981ApJ...245L..79C}).  More
   recent works confirm star-to-star light-element variations in NGC\,6752
   (\citealt{2002A&A...385L..14G}; \citealt{2003A&A...402..985Y,
     2008ApJ...684.1159Y,2013MNRAS.434.3542Y};
   \citealt{2005A&A...433..597C}), O-Na, Mg-Al, C-N (anti-)correlations
   for both unevolved (\citealt{2001A&A...369...87G};
   \citealt{2010A&A...524L...2S}; \citealt{2010A&A...524A..44P}) and RGB
   stars (\citealt{2005A&A...438..875Y};
   \citealt{2007A&A...464..927C,2012ApJ...750L..14C}) in close analogy
   with what was observed in most Galactic GCs (see e.g.\,
   \citealt{2002AJ....123.3277R}, \citealt{2009A&A...505..117C} and
   references therein). In particular, there are three main groups of
   stars with different Na, O, N, Al, that populates three different RGBs
   when appropriate indices are used (like the $c_{1}$ and $c_{\rm y}$
   Stromgren indices or the $c_{U,B,I}$ visual index,
   \citealt{2008ApJ...684.1159Y, 2003A&A...402..985Y},
   \citealt{2012ApJ...750L..14C}, \citealt{2013MNRAS.431.2126M}).

   As shown by \citealt{2013ApJ...767..120M} (hereafter Mi13), the CMD
   of NGC\,6752 is made of three distinct sequences that can be
   followed continuously from the MS to the SGB and from the SGB to
   the RGB. These sequences correspond to three stellar populations
   with different light-element and helium abundance.

   \section{Observations and data reduction \label{obs}}
   
   \begin{figure}
     \centering
     \includegraphics[bb=34 323 517 680, width=\hsize]{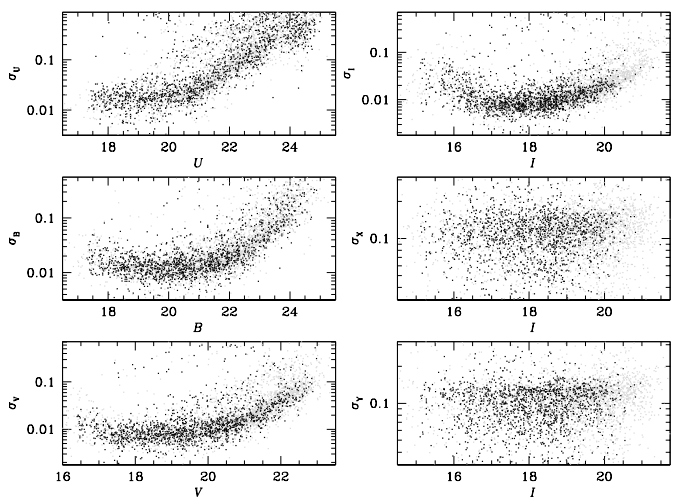} 
     \caption{ The photometric (left panels and top-right
         panel) and position residuals (middle- and bottom-right
         panels) from the single measurements in the single images
       of NGC\,6121 plotted as a function of the average
       magnitude. Grey points show all detected stars; black points
       refers to proper motion selected stars.  In the case of
         NGC\,6752 and NGC\,6397 the distributions are similar. }
     \label{rms}
   \end{figure}

   For this work we used $2 \times 2$ binned images taken with the
   ESO/FORS2 ($2\times 2\mathrm{k}\times 4\mathrm{k}$ MIT CCDs)
   mounted at the VLT, using the standard resolution collimator. With
   this configuration, the field of view of FORS2 is reduced to $\sim
   6\farcm 8 \times 6\farcm 8$ by the MOS unit in the focal plane and
   the pixel scale is $\sim 0\farcs 25/\mathrm{pixel}$. The dithered
   images of NGC\,6121, NGC\,6397 and NGC\,6752 were acquired using
   u\_HIGH, b\_HIGH, v\_HIGH and I\_Bessel broad band filters between
   April 14, 2012 and July 23, 2012. A detailed log of observations is
   reported in Table \ref{table:1}. Figure \ref{footprint} shows the
     combined field of view for each cluster.

   For the data reduction we used a modified version of the software
   described in \cite{2006A&A...454.1029A}. Briefly, for each image we
   obtained a grid of 18 spatially varying empirical point spread
   functions (PSFs, an array of $3 \times 3$ PSFs for each chip of
   FORS2) using the most isolated, bright and not saturated stars. In
   this way, to each pixel of the image corresponds a PSF that is a
   bi-linear interpolation of the closest four PSFs of the grid. This
   makes it possible to measure star positions and fluxes in each
   individual exposure using an appropriate PSF and to obtain a
   catalog of stars for each frame. We registered, for each cluster
   and for each filter, all star positions and magnitudes of each
   catalog into a common frame (master-frame) using linear
   transformations. The final result is a list of stars (master list)
   for each cluster.  For each filter and each star measured in NGC
   6121, in Fig.~\ref{rms} we plot the rms of the photometric residual
   and of the position as a function of the mean magnitude.  In the
   case of NGC 6752 and NGC 6397 the distributions are similar.  
 As required by the referee, we specify here that we used magnitude to
 express the luminosities of stars in this paper.

\begin{table}
  \caption{Log of observations}
  \label{table:1}  
  \centering       
  \begin{tabular}{l c c c}
    \hline\hline              
    \textbf{Filter} & \textbf{Exp. time } & \textbf{Airmass} & \textbf{Seeing} \\
    \, & & ($\sec{z}$) & (arcsec) \\
    \hline                     
    \multicolumn{4}{c}{\textbf{NGC\,6121}} \\
    u\_HIGH & $25 \times 410$\,s  & 1.004--1.104 &  $0\farcs58$--$0\farcs90$ \\
    b\_HIGH & $25 \times 200$\,s  & 1.007--1.113 &  $0\farcs71$--$1\farcs20$ \\
    v\_HIGH & $25 \times \,\,52 $\,s  & 1.118--1.251 &  $0\farcs70$--$1\farcs21$ \\
    I\_BESS & $25 \times \,\,30 $\,s  & 1.036--1.150 &  $0\farcs56$--$1\farcs03$ \\
    \multicolumn{4}{c}{\textbf{NGC\,6397}}\\
    u\_HIGH & $25 \times 410$\,s  & 1.153--1.452 &  $0\farcs81$--$1\farcs09$ \\
    b\_HIGH & $35 \times 200$\,s  & 1.144--1.599 &  $0\farcs73$--$1\farcs52$ \\
    v\_HIGH & $50 \times 52\,\, $\,s  & 1.142--1.272 &  $0\farcs82$--$1\farcs23$ \\
    I\_BESS & $25 \times 30\,\, $\,s  & 1.251--1.355 &  $0\farcs70$--$1\farcs21$ \\
    \multicolumn{4}{c}{\textbf{NGC\,6752}}\\
    u\_HIGH & $25 \times 410$\,s  & 1.227--1.416 &  $0\farcs67$--$1\farcs30$ \\
    b\_HIGH & $33 \times 200$\,s  & 1.227--1.850 &  $0\farcs49$--$1\farcs36$ \\
    v\_HIGH & $25 \times 52\,\, $\,s  & 1.367--1.483 &  $0\farcs68$--$0\farcs86$ \\
    I\_BESS & $25 \times 30\,\, $\,s  & 1.293--1.363 &  $0\farcs50$--$0\farcs77$ \\
    \hline
  \end{tabular}
\end{table}

\begin{figure*}
  \centering
  \includegraphics[bb=35 290 580 670, width=0.95\hsize]{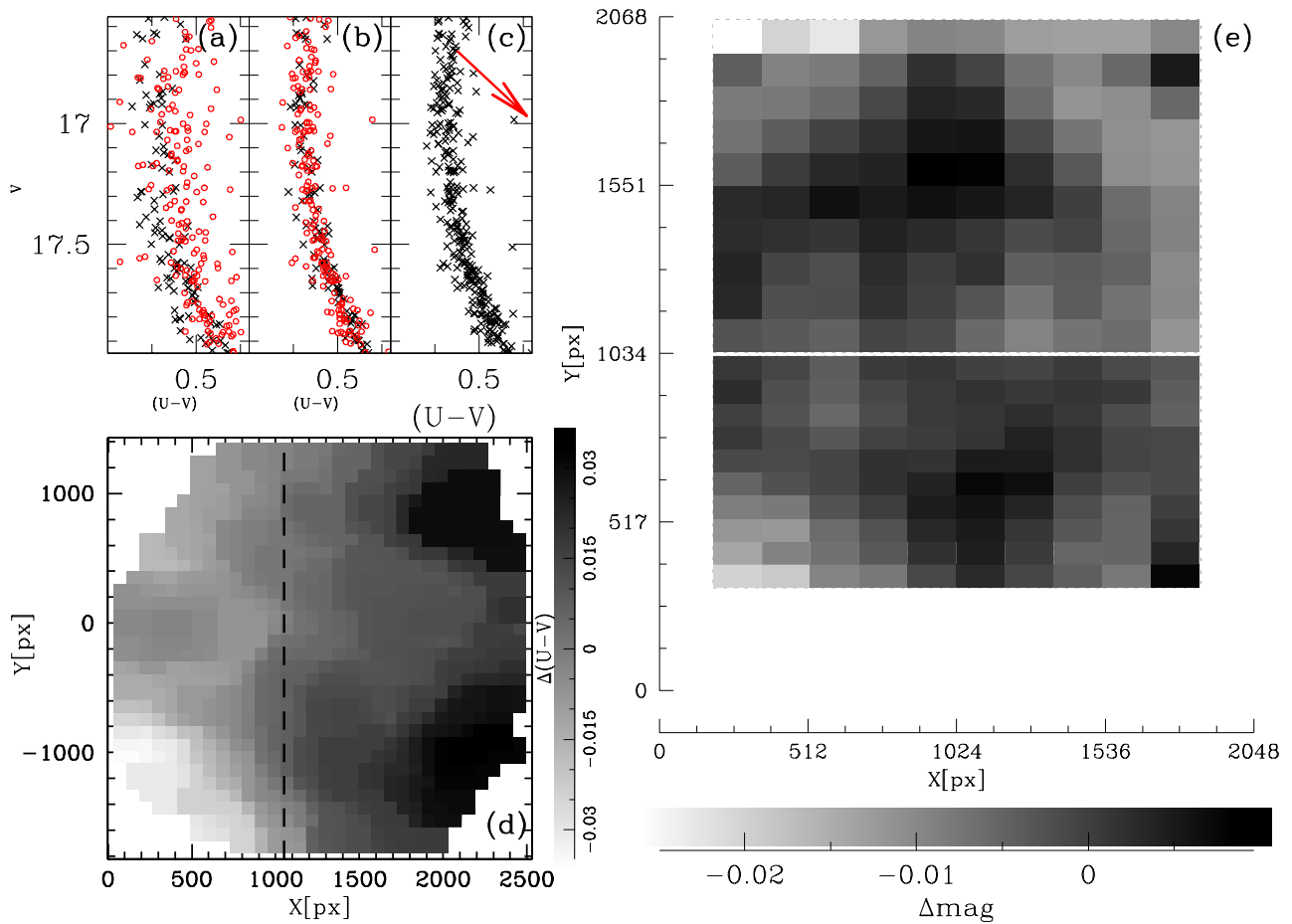}
  \caption{Visualization of the effects of the photometric zero point
    variation. Panel (a): CMDs zoomed into the MS region of NGC\,6397
    before the zero-point correction; panel (b): CMD after the
    zero-point correction; panel (c): CMD after the differential
    reddening correction (in red the reddening vector, scaled by a
    factor 1/3) ; panel (d) map of the color zero point variation; the
    dashed line divide the two groups of stars showed in the panels
    (a) and (b); panel (e): final $\Delta$mag correction grid for the
    filter v\_HIGH and the NGC\,6397 data-set. }
  \label{zp}
\end{figure*}

We noted that all the CMDs of the three GCs showed unusually spread
out sequences (see panel (a) of Fig. \ref{zp} for an example of the
$V$ versus $U-V$ CMD of NGC\,6397).  Part of this spread is due to
differential reddening, but this is not the only cause.  In fact, we
found that, selecting stars in different regions of the master list,
we obtained shifted MSs. As an example, in panel (d) of Fig. \ref{zp}
we show the variation of the color $\Delta(U-V)$ for NGC\,6397. This
plot shows that there is an important gradient of $\Delta(U-V)$ along
the x-axis. We selected in this plot two groups of stars: the stars
with $x>1050$ and that with $x<1050$. We plotted these two sub-samples
in the $V$ versus $U-V$ CMDs, respectively in black crosses ($x>1050$) and red circles
($x<1050$): panel (a) of Fig. \ref{zp} shows the result. The two
groups form two shifted MSs. This effect is present in all the CMD of
the three GCs, even if with different extent levels.

\citet{2007ASPC..364..113F} showed that there is an illumination
gradient in the FORS2 flats produced by the twilight sky. This
gradient changes with time and with the position of the Sun relative
to the pointing of the telescope and could produce a photometric zero
point variation across the FORS2 detectors.

The illumination gradient in the flat fields could contribute to the
observed enlargement of the CMDs. 

We obtained star fluxes using local sky values, and therefore it is
expected that these systematic effects are negligible. If the gradient
in the flat-field images is not properly removed during the
pre-reduction procedure, the pixel quantum efficiency correction will
be wrong. The consequence is that the luminosity of a star measured in
a given location of the CCD will be underestimated (or overestimated)
with respect to the luminosity of the same star measured in another
location of the CCD.

Using the measured star positions and fluxes, we performed a
correction in a similar way to what described by
\citet{2009A&A...493..959B}. It is a self-consistent auto-calibration
of the illumination map and takes advantage of the fact that the
images are well dithered.

For each cluster and for each filter, the best image 
     (characterized by lower airmass and best seeing) is defined as
reference frame. We considered the measured raw magnitude $m_{ij}$ of
each star $i$ in each image $j$. Using common stars between the image
$j$ and the reference frame we computed the average magnitude shift:
\[
\Delta_j=\frac{1}{M}\sum_{i=1}^M (m_{ij}-m_{i,\mathrm{ref}}) 
\]
where M is the number of stars in common between the reference frame
and the single image $j$. For each star, that is centered in a
different pixel in each dithered frame, we computed its average
magnitude in the reference system:
\[
{\bar m} = \frac{1}{N} \sum_{j=1}^N (m_{ij}-\Delta_{j})
\]
where $N$ is the number of images in which the star appears. Then we
computed the residual:
\[
\delta_{ij} = m_{ij}-\Delta_{j} - {\bar m_{j}}\,.
\]
We divided each FORS2 chip in a spatial grid of $10 \times 10$ boxes,
and, for each box, we computed the average of the residuals from the
stars located in that region in each single image. This provides a
first spatial correction to our photometry. To obtain the best
correction, we iterate until the residual average became smaller than
1 mmag. To guarantee convergence we applied, for each star, half of
the correction calculated in each box. Moreover, to obtain the best
correction at any location of the camera, we computed a bi-linear
interpolation of the closest 4 grid points. At the edges of the
detectors the correction is less efficient, because the corresponding
grid-points have been moved toward the external borders of the grid to
allow the bi-linear interpolation to be computed all across the
CCDs. In panel (e) of Fig.\,\ref{zp} we show our final correction grid
for the v\_HIGH filter.

The final correction grids are different for each filter and for each
set of images. In particular, the patterns are different from filter
to filter, as well as the size of the zero point variations. This
quantity also changes using different data-sets.  The maximum
amplitudes of our corrections are tabulated in Table 2.

   \begin{table}
     \caption{Maximum amplitudes of photometric zero-points corrections.}
     \label{table:2b}  
     \centering       
     \begin{tabular}{l c c c}
       \hline\hline              
       \textbf{Filter} & \multicolumn{3}{c}{\textbf{$\Delta$mag}} \\
       \hline
       \textbf{} & \textbf{NGC\,6121 } & \textbf{NGC\,6397} & \textbf{NGC\,6752} \\
       \hline                     
       u\_HIGH & 0.11 & 0.13 & 0.09 \\
       b\_HIGH & 0.04 & 0.05 & 0.03 \\
       v\_HIGH & 0.12 & 0.04 & 0.06 \\
       I\_BESS & 0.04 & 0.04 & 0.04 \\
       \hline
     \end{tabular}
   \end{table}
   \begin{figure*}
     \centering
     \includegraphics[bb= 85 245 518 590, width=\hsize]{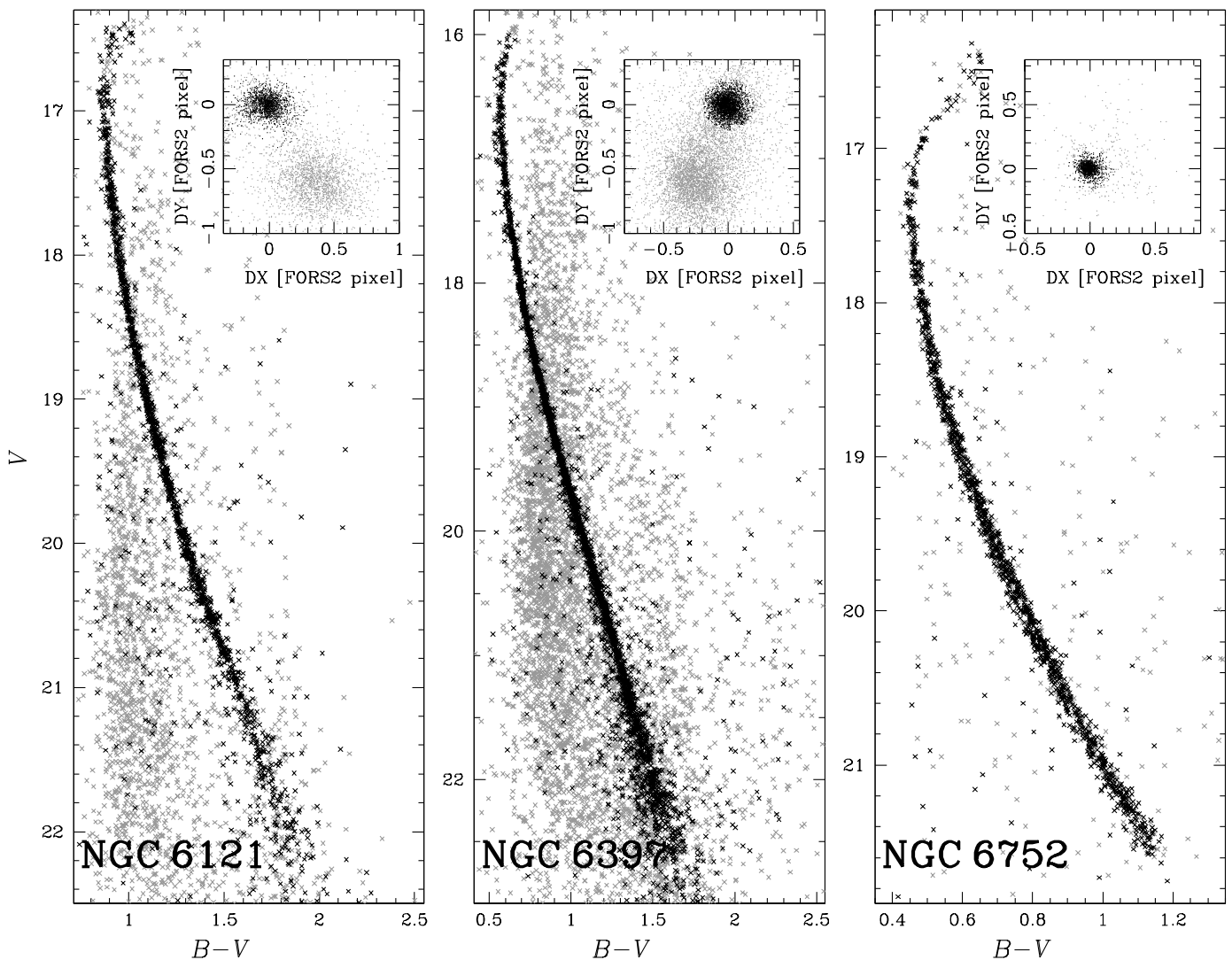} 
     \caption{$V$ versus $B-V$ CMD of stars in the field of view of
       NGC\,6121 (left), NGC\,6397 (middle), and NGC\,6752
         (right). The insets show the vector-point diagram of stellar
       displacements along the X and Y direction. Black and gray
       points indicate stars that, according to their proper motions,
       are considered cluster members and field stars, respectively
       (see text for details).}
     \label{pm}
   \end{figure*}

We corrected the spread of the CMD due to differential reddening using
the procedure as described by \citet{2012A&A...540A..16M}. Briefly, we
defined a fiducial line for the MS of the cluster. Then, for each
star, we considered a set of neighbors (usually 30, selected anew for
each filter combination) and estimated the median offset relative to
the fiducial sequence. These systematic color and magnitude offsets,
measured along the reddening line, represent an estimate of the local
differential reddening. With this procedure we also mitigated the
photometric zero-point residuals left by the illumination correction
(especially close to the corners of the field of view).  Panel (c) of
Fig. \ref{zp} shows the CMD after all the corrections are applied.

The photometric calibration of FORS2 data for $UBV$ Johnson and $I_C$
Cousins bands was obtained using the photometric Secondary Standards
star catalog by \citet{2000PASP..112..925S}. We matched our final
catalogs to the Stetson standard ones, and derived calibration
equations by means of least squares fitting of straight lines using
magnitudes and colors.

\subsection{Proper motions}

Since NGC\,6121 (l,b=350\fdg97,15\fdg97) and NGC\,6397 (l,b=338\fdg17;$-$11\fdg96)
are projected at low Galactic latitude, their CMDs are both
dramatically contaminated by Disk and Bulge stars, contrary to
NGC\,6752 (l,b=336\fdg49;$-$25\fdg63) that presents low field contamination.
The average proper motions of NGC\,6121, NGC\,6397 and NGC\,6752
strongly differ from that of these field stars
(e.g.\,\citealt{2003AJ....126..247B},
\citealt{2006A&A...456..517M}). Therefore, to minimize the
contamination from field stars we identified cluster members on the
basis of stellar proper motions.

In order to get information on the cluster membership, we estimated
the displacement between the stellar positions obtained from our FORS2
data-set and those in the ground-based data taken from the image
archive maintained by \citet{2000PASP..112..925S} and also used in
\citet{2013MNRAS.431.2126M}. These observations include images from
different observing runs with the Max Planck 2.2m telescope, the CTIO
4m, 1.5m and 0.9m telescopes and the Dutch 0.9m telescope on La
Silla. To obtain the displacement, we used six-parameter local
transformations based on a sample of likely cluster members, in close
analogy to what was done by \citet{2003AJ....126..247B} and
\citet{2006A&A...454.1029A} to calculate stellar proper motions.

Results are shown in Fig.~\ref{pm}. The figure shows the $V$ versus
$B-V$ CMDs for NGC\,6121, NGC\,6397 and NGC\,6752. The insets show the
vector-point diagrams of the stellar displacements for the same stars
shown in the CMDs: the cluster-field separation is evident. 
  Likely cluster members are plotted in black both in the CMDs and in
  the vector-point diagrams, in gray the rejected stars.

   \begin{figure*}
     \centering
     \includegraphics[bb=60 375 520 660, width=0.9\hsize]{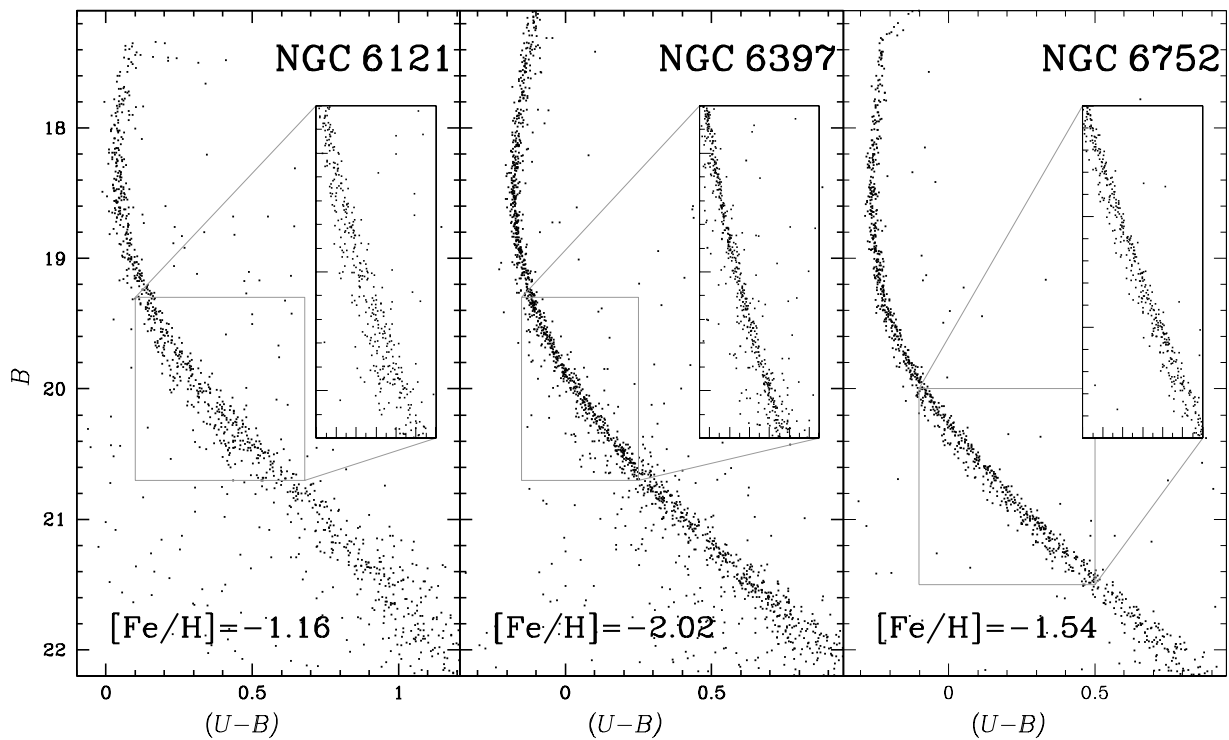}
     \caption{$B$ versus $(U-B)$ CMD for NGC\,6121 (left), NGC\,6397
       (middle), and NGC\,6752 (right). The inset is a zoom around the
       upper MS. The CMDs only show the cluster members and are
       corrected by zero point variations and differential reddening}
     \label{ub}
   \end{figure*}

\section{The CMDs of the three GCs \label{CMDs}}

Previous studies on multiple stellar populations have demonstrated
that the $U-B$ color is very efficient in detecting multiple RGBs (see
\citealt{2008A&A...490..625M} and \citealt{2010ApJ...709.1183M} for the
cases of NGC\,6121 and NGC\,6752), and multiple MSs (see
\citealt{2012A&A...540A..16M}, Mi13 for the cases of NGC\,6397 and
NGC\,6752).
As discussed by \citet{2011A&A...534A...9S}, CNO abundance variations
affect wavelengths shorter than $\sim 400$\,nm owing to the rise of
molecular absorption bands in cooler atmospheres. The consequences are
that the CMDs in $UB$ filters show enlarged sequences, mainly due
to variations in the N abundance, with the largest
variations affecting the RGB and the lower MS.

Motivated by these results, we started our analysis from the $B$ versus
$U-B$ CMDs shown in Fig.~\ref{ub}; the inset of each panel is a zoom in
of the upper MS, between $\sim 1.5$ and $\sim 2.5$ magnitudes below
the turn off. 

A visual inspection at these CMDs reveals that the color broadening of
MS stars in both NGC\,6121 and NGC\,6752 is larger than that of
NGC\,6397. A small fraction of MS stars in NGC\,6121 and NGC\,6752
defines an additional sequence on the blue side of the most-populous MS.

\begin{figure*}
  \centering
  \includegraphics[bb= 26 67 555 590,width=1\hsize]{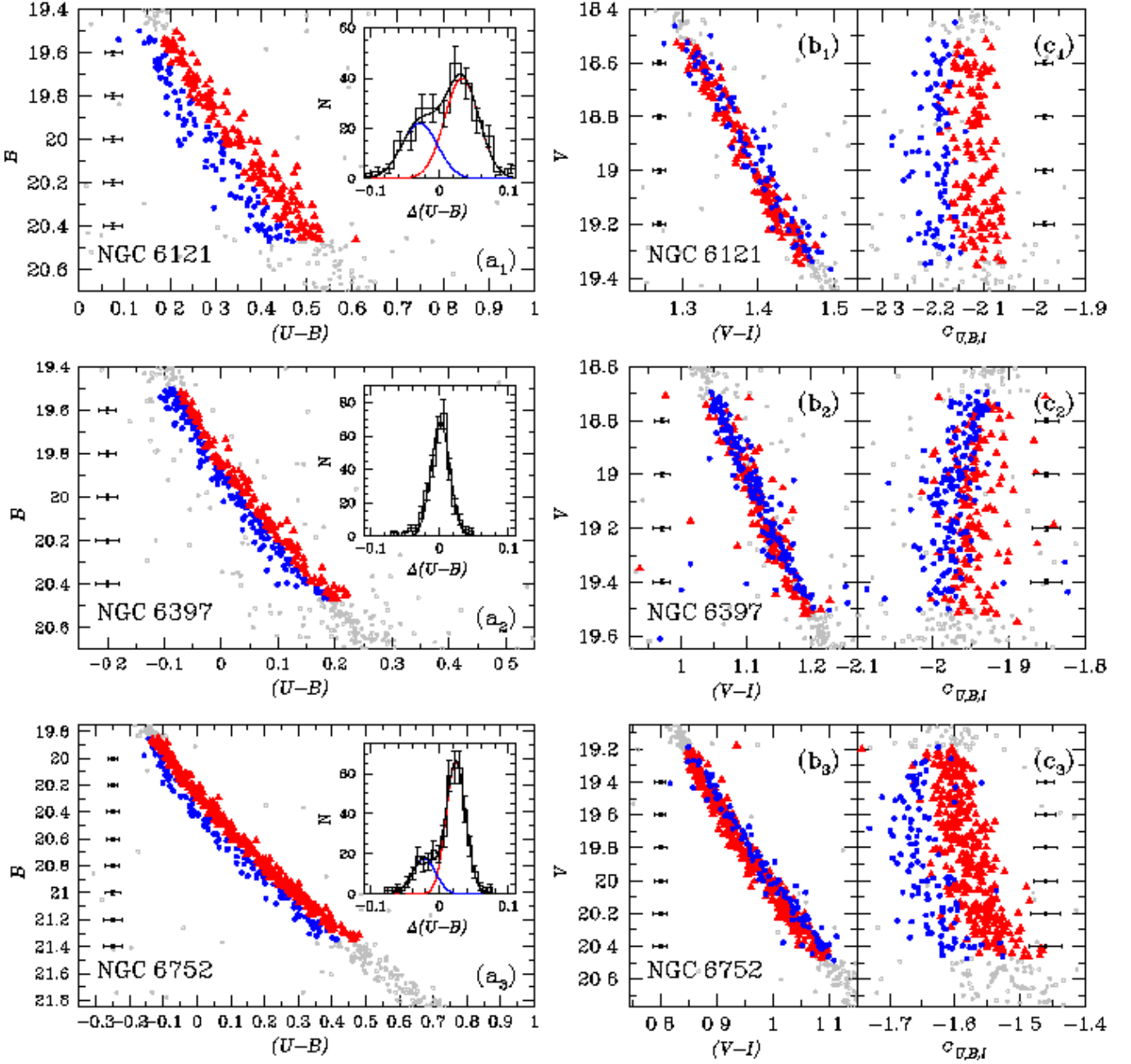}
  \caption{Comparison between the $B$ vs. $U-B$ (panels a), the $V$
    vs. $V-I$ (panels b) and $V$ vs $c_{U,B,I}$ (panels c) CMDs of
    NGC\,6121 (up), NGC\,6397 (middle) and NGC\,6752 (down). Red 
      triangles  and blue points represent the groups of rMS and bMS
    stars defined in panels (a). The insets of panels (a) show the
    $\Delta (U-B)$ distributions, fitted with a bi-Gaussian in the
    case of NGC\,6121 and NGC\,6752 and with a Gaussian in the case of
    NGC\,6397.  The horizontal bars show the mean error in color.}
  \label{prove}
\end{figure*}
\begin{figure*}
  \centering
  \includegraphics[bb=22 479 580 677, width=1\hsize]{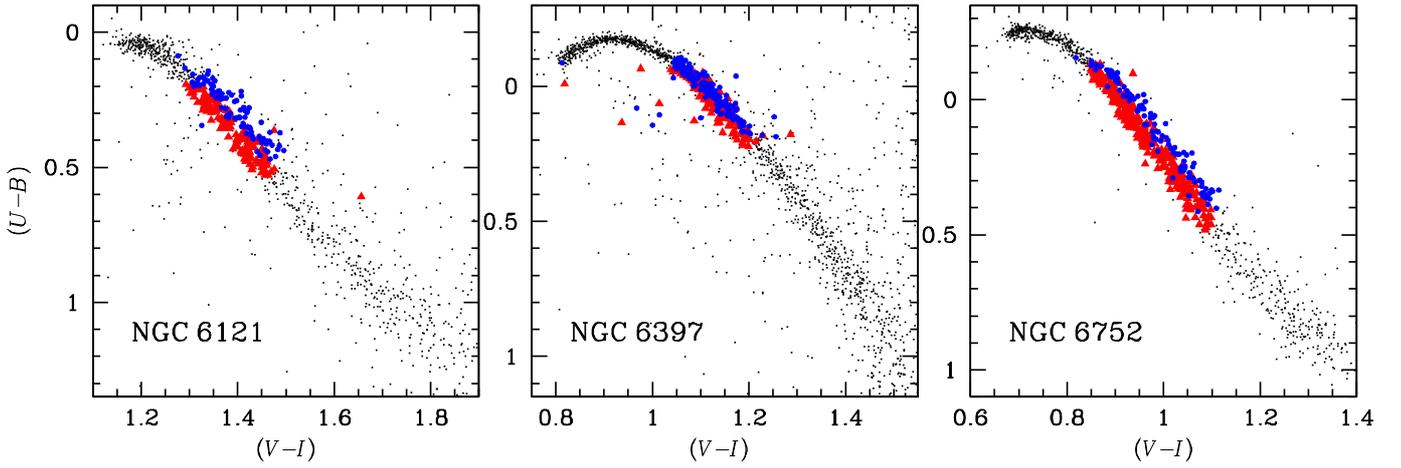}
  \caption{Color-color diagrams for NGC\,6121 ({\it left panel}),
    NGC\,6397 ({\it central panel}) and NGC\,6752 ({\it right
      panel}). In red  (triangles) and blue dots the rMS
    and bMS as defined in Fig.~\ref{prove}}
  \label{colcol}
\end{figure*}
\begin{figure*}
  \centering
  \includegraphics[bb=122 311 470 660, width=0.7\hsize]{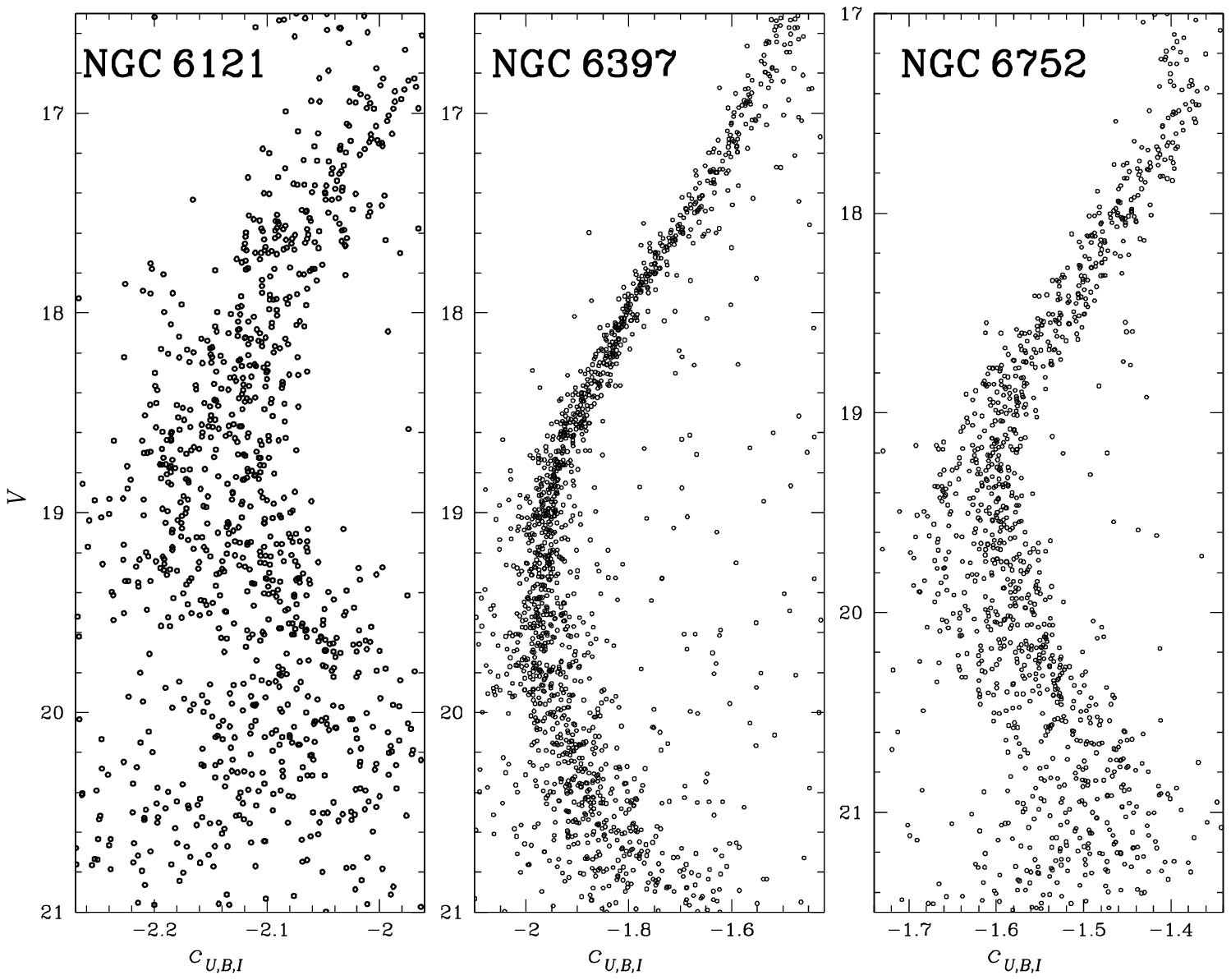}
  \caption{$V$, $c_{ U,B,I}$ diagrams for NGC\,6121 (left),
      NGC\,6397 (middle) and NGC\,6752 (right). }
  \label{cubi}
\end{figure*}

To investigate whether the widening of the MSs is due to the presence
of multiple populations we identified in the $B$ versus $U-B$ CMD of
each cluster two groups of red-MS (rMS) and blue-MS (bMS) stars, as
shown in panels (a) of Fig.~\ref{prove}. We colored the two MSs in red
and blue respectively, and these colors are used consistently
hereafter. In the case of NGC\,6121 and NGC\,6752, where there is some
hint of a split MS, we identified by eye the fiducial that divide the
rMS and bMS. In the case of NGC\,6397 we have considered as rMS (or
  bMS), the stars that are redder (or bluer) than the MS fiducial
  line. Each inset shows the color distribution of $\Delta(U-B)$ for
  the two MSs, where $\Delta(U-B)$ is obtained by subtracting the
  color of the fiducial that divide the two MSs to the color of the
  rMS and bMS stars. In the cases of NGC\,6121 and NGC\,6752 the color
  distributions show a double peak that could be due to the presence
  of two populations; we fitted them with a sum of Gaussians (in red
  and blue respectively for the rMS and bMS).  We applied a moving box
  procedure to further verify that the distribution of $\Delta(U-B)$ in
  the case of NGC6121 is bimodal. We changed the binsize, ranging from
  0.005 to 0.02 (approximately the error in color) with steps of
  0.001.  Furthermore, from our dataset we determined a kernel-density
  distribution by assuming a Gaussian kernel with $\sigma=$0.02 mag.
  I all cases, we consistently found that the distribution can only be
  reproduced by two Gaussians.

A multiple sequence in NGC 6752 had already been identified by Mi13
using {\it HST} data. Very recently, we had the first F275W, F336W,
F438W WFC3 images of NGC\,6121 from the {\it HST} GO-13297 UV Large
Legacy Program (P.I.~Piotto). Even a preliminary reduction of the data
shows a clear separation of the MS into two branches in the F438W vs
F336W-F438W CMD, fully confirming what we anticipate here in the
equivalent, groundbased $U$ vs $U-B$ diagram of Fig.~\ref{prove}.

In the case of NGC\,6397, it is possible to fit the distribution with
a single Gaussian.

As an additional check for the presence of multiple populations, we
investigated whether the widening of the MSs of all the GCs is
intrinsic or if it is entirely due to photometric errors.  We have
compared two CMDs, $B$ versus $U-B$ with $V$ versus $V-I$, obtained
using independent datasets. We considered the rMS and bMS defined
previously: if the color spread is entirely due to photometric errors,
a star which is red (or blue) in the $B$ versus $U-B$ CMD will have
the same chance of being either red or blue in the $V$ vs $V-I$ . By
contrast, the fact that the two sequences identified in the first CMD
have systematically different colors in the second one, would be a
proof that the color broadening of the MS is intrinsic. In panels (b)
of Fig.~\ref{prove} we plotted the rMS and bMS in $V$ versus $V-I$
CMDs. The fact that the rMS stars of both NGC\,6121 and NGC\,6752
have, on average, different $V-I$ than bMS stars demonstrates that the
color broadening of their MS in the $U$ versus $U-B$ CMDs is
intrinsic. This is the first evidence that the MS of NGC\,6121 is not
consistent with a simple stellar population. In the case of NGC\,6397
rMS and bMS stars share almost the same $V-I$ thus suggesting that
most of the colors broadening is due to photometric errors.

As last test, we plotted the two MSs in the
$V$ versus $c_{U,B,I}$ CMD. The $c_{U,B,I}$ index, which is defined as
the color difference $(U-B)-(B-I)$, is a very efficient tool to
identify multiple sequences in GCs. Indeed it maximizes the color
separation between the stellar populations that is due to both helium
and light-element variations (\citealt{2013MNRAS.431.2126M}). Panels
(c) of Fig.~\ref{prove} confirm the previous results: rMS and bMS of
NGC\,6121 and NGC\,6752 are well defined in the $V$ versus $c_{U,B,I}$
CMDs, but this is less evident for NGC\,6397.

As a last prof, we plotted in Fig.~\ref{colcol} the $(U-B)$ versus
$(V-I)$ diagrams for each cluster: in red and in blue are plotted the
rMS and bMS defined previously. The figure shows that for both
NGC\,6121 and NGC\,6752 the two MSs are well defined, while for
NGC\,6397 the rMS and bMS stars are mixed

\begin{figure}
     \centering
     \includegraphics[bb=60 456 285 670, width=\hsize]{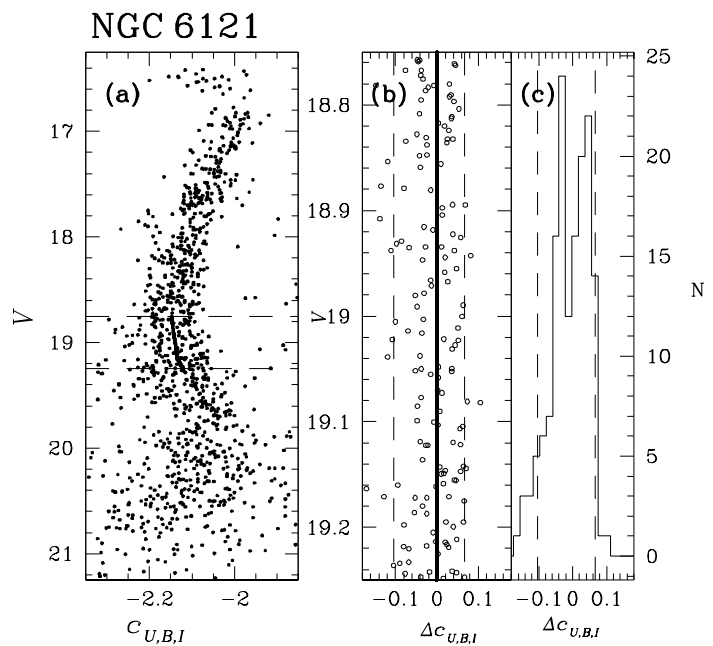} \\
     \caption{Procedure to estimate the MS width for NGC\,6121.
       Panel (a) show the $V$ versus $c_{U,B,I}$ CMD of NGC\,6121. The
       thick line is the MS fiducial line (see text for details). Panel (b)
       shows the verticalized MS between the two dashed lines of panel
       (a). Panel (c) is the distribution of the color for the stars of
       panel (b).}
     \label{dcubi_histo}
   \end{figure}

Figure~\ref{cubi} shows the $V$ versus $c_{U,B,I}$ diagram for the
three GCs studied in this paper. In their analysis of multiple stellar
populations in 22 GCs, \citet{2013MNRAS.431.2126M} found that all the
analyzed clusters show a multimodal or spread RGB in the $V$ versus
$c_{U,B,I}$ diagram, and the $c_{U,B,I}$ value of each star depends on
its light-element abundance. The $c_{U,B,I}$-index width of the RGB
($W_{\rm RGB}$) correlates with the cluster metallicity, with the more
metal rich GCs having also the largest values of $W_{\rm RGB}$. In
order to compare the MSs of the three GCs studied in this paper, we
introduce a quantity, $W_{\rm MS}$, which is akin of $W_{\rm RGB}$,
but is indicative of the $c_{ U,B,I}$-index broadening of the MS. The
procedure to determine $W_{\rm MS}$ is illustrated in
Fig.~\ref{dcubi_histo} for NGC\,6121 and is the same for the other
clusters.  We have considered the magnitude of the TOs as magnitude of
reference: $V_{\rm MSTO} \sim 16.6$ in the case of NGC\,6397, $V_{\rm
  MSTO} \sim 17.4$ for NGC\,6752, and $V_{\rm MSTO} \sim 16.75$ for
NGC\,6121.  Panel (a) of Fig.~\ref{dcubi_histo} shows the $V$ versus
$c_{ U,B,I}$ diagram for NGC\,6121 in a range of magnitudes from
$V_{MSTO}-0.5$ to $V_{MSTO}+5$ . In this range of magnitudes, we
obtained the fiducial line for the MSs computing the $3.5
\sigma$-clipped median of the color in interval of 0.35 mag and
interpolated these points with a spline. In our analysis we used only
MS stars with $2<V-V_{\rm MSTO}<2.5$ where the MS split is visible for
both NGC\,6121 and NGC\,6752. This magnitude interval is delimited by
the two dashed lines of Fig.~\ref{dcubi_histo}a. The thick line is the
fiducial in the considered magnitude interval. The verticalized $V$
versus $\Delta c_{ U,B,I}$ diagrams is plotted in panel (b) of
Fig.~\ref{dcubi_histo}, while panel (c) shows the histogram
distribution of $\Delta c_{ U,B,I}$.  The MS width, $W_{\rm MS}$, is
defined as the $\Delta c_{ U,B,I}$ extension of the histogram and is
obtained by rejecting the 5\% of the reddest and the bluest stars on
the extreme sides.  To account for photometric error, we have
subtracted from the observed $W_{\rm MS, OBS}$ the average error in
$c_{U,B,I}$ in the same magnitude interval, i.e. $W_{\rm
  MS}=\sqrt{W_{\rm MS, OBS}^2-\sigma_{c_{U,B,I}}^2}$.

We found that the most metal-rich GC, NGC\,6121, exhibits the largest
$c_{ U,B,I}$ index width for MS stars ($W_{\rm
  MS}=0.169\pm0.014$).\\ The spread in $c_{U,B,I}$ is smaller in the
case of NGC\,6752 ($W_{\rm MS}=0.115\pm0.006$) and drops down to
$W_{\rm MS}=0.093\pm0.014$ in the most metal-poor GC NGC\,6397. To
estimate the statistical uncertainty in measuring $W_{\rm MS}$, we
used the bootstrap resampling of the data to generate 10000 samples
drawn from the original data sets. We computed the standard deviation
from the mean of the simulated $W_{\rm MS}$ and adopted this as
uncertainty of the observed $W_{\rm MS}$.

These findings make it tempting to speculate that the $c_{ U,B,I}$
index width of the MS could be correlated with the cluster
metallicity, in close analogy with what observed for RGB stars. An
analysis of a large sample of GCs is mandatory to infer any conclusion
on the relation between $W_{\rm MS}$ and [Fe/H].

Mi13 have identified three stellar populations in NGC\,6752 that they
have named as `a', `b', and `c'. Population `a' has a
chemical composition similar to field halo stars of the same
metallicity, population `c' is enhanced in sodium and nitrogen,
depleted in carbon and oxygen and enhanced in helium ($\Delta Y
\sim$0.03), while population `b' has an intermediate chemical
composition between `a' and `c' and is slightly helium enhanced
($\Delta Y \sim$0.01).  However, the MSs of populations `b' and `c'
are nearly coincident in the $m_{\rm F336W}$ versus $m_{\rm
  F336W}-m_{\rm F390W}$ CMD, while population `a' stars have bluer
$m_{\rm F336W}-m_{\rm F390W}$ colors (Mi13, see their Fig.~8). The
three MSs exhibit a similar behavior also in the $m_{\rm F336W}-m_{\rm
  F410M}$ and $m_{\rm F336W}-m_{\rm F467M}$ colors. Since these colors
are similar to $U-B$, the less populous MS identified in this paper
should correspond to the population `a' identified by Mi13, while the rMS
hosts both population `b' and population `c' stars.

\subsection{The fraction of rMS and bMS in NGC\,6121 and NGC\,6752} \label{popratio}

\begin{figure*}
  \centering
  \includegraphics[bb=20 230 570 545, width=\hsize]{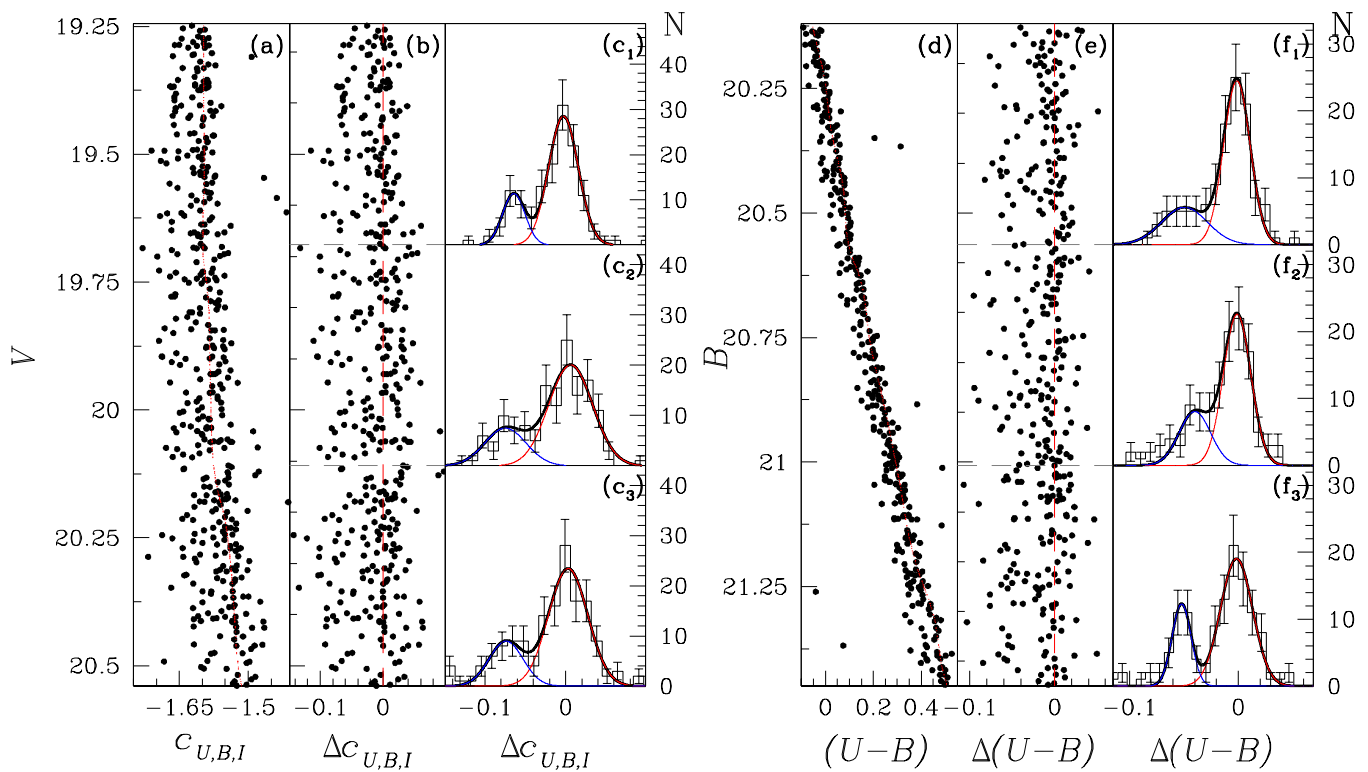}
  \caption{ Procedure to estimate the fraction of rMS and bMS stars in
    NGC~6752 by using the $V$ versus $c_{U,B,I}$ diagram (panels (a),
    (b) and (c)), and the $B$ versus $U-B$ CMD (panels (d), (e) and
    (f)).  Panels (a) and (d) reproduce the same diagrams as
    Figs.~\ref{cubi} and~\ref{ub}. The red line is the rMS fiducial
    line. Panels (b) and (e) show the verticalized MS. The histogram
    distribution of $\Delta({\rm color})$ for the stars of the panel
    (b) and (e) is plotted in panels (c) and (f) for three intervals
    of magnitude. The tick black lines, superimposed to each
    histogram, are the best-fitting biGaussian functions, whose
    components are colored red and blue. In the case of $V$
      versus $c_{U,B,I}$ CMD, we obtained that rMS and bMS contain
      respectively $77\pm4$\,\% and $23\pm8$\,\% of MS stars in panel
      (c1), $74\pm5$\,\% and $26\pm8$\,\% of MS stars in panel (c2)
      and, $75\pm5$\,\% and $25\pm8$\,\% of the MS stars. In the case
      of $B$ versus $(U-B)$ CMD we obtained that rMS a bMS contains
      respectively $72\pm5$\,\% and $28\pm8$\,\% of MS stars in panel
      (f1), $74\pm6$\,\% and $26\pm9$\,\% of MS stars in panel (f2),
      $73\pm5$\,\% and $27\pm8$\,\% of MS stars in panel (f3).}
  \label{n6752}
\end{figure*}

In order to measure the fraction of stars in each MS we followed the
procedure illustrated in Fig.~\ref{n6752} for NGC\,6752, which we
already used in several previous papers
(e.g.\,\citealt{2007ApJ...661L..53P}, Mi13).

Panel (a) shows the $V$ versus $c_{U,B,I}$ CMD of the MS stars in the
magnitude interval $19.25<V<20.55$, where the MS split is most
evident. We verticalized the selected MS by subtracting the color of
the stars to the color of the fiducial line of the rMS, obtaining
$\Delta c_{U,B,I}$. The fiducial line is obtained by hand selecting
the stars of the rMS, dividing them in bins of magnitude, computing
the median colors of the stars within each bin and interpolating these
median points with a spline. The verticalized $V$ versus $\Delta c_{
  U,B,I}$ diagram is plotted in panel (b).

The $\Delta c_{U,B,I}$ color distribution of the stars for three
magnitude intervals is shown in panels (c). Each histogram clearly
shows two peaks and has been simultaneously fitted with a double
Gaussian, whose single components are shown in blue and in red for the
bMS and the rMS, respectively.

For each magnitude interval, from the area under the Gaussians we
infer the fraction of bMS and rMS stars. The errors ($\sigma$)
associated to the fraction of stars are estimated as $\sigma = \sqrt
{\sigma_{\rm I}^{2} + \sigma_{\rm II}^{2}}$, where $\sigma_{\rm I}$ is
derived from binomial statistics and $\sigma_{\rm II}$ is the
uncertainty introduced by the histogram binning and is derived as in
\citet{2014A&A...563A..80L}.  Briefly, we have derived $N$ times the
population ratio as described above, but by varying the binning
  and starting/ending point in the histogram. We assumed as
$\sigma_{\rm II}$ the rms scatter of these $N$ determinations.

We computed the weighted mean of the bMS and rMS fractions of the
three magnitude intervals, using as weight $w=1/\sigma^2$. In the case
of $V$ versus $c_{U,B,I}$, we obtained that the rMS and bMS contain
respectively $75\%\pm3\%$ and $25\%\pm5\%$ of MS stars. In panels (d),
(e) and (f) of Fig.~\ref{n6752} we applied the same procedure in the
$B$ versus $(U-B)$ using the stars with $20.10<B<21.45$.  We obtained
that the blue MS contains $27\%\pm 5\%$ of the total number of MS
stars, and the red MS is made of the remaining $73\% \pm 3\%$ stars.
We also calculated the weighted mean of the results of the two CMDs of
NGC\,6752, obtaining that in the rMS there are the $74\%\pm 2\%$ of
stars, and in the bMS the remaining $26\%\pm4\%$.

\begin{figure*}
  \centering
  \includegraphics[bb=20 230 570 545, width=\hsize]{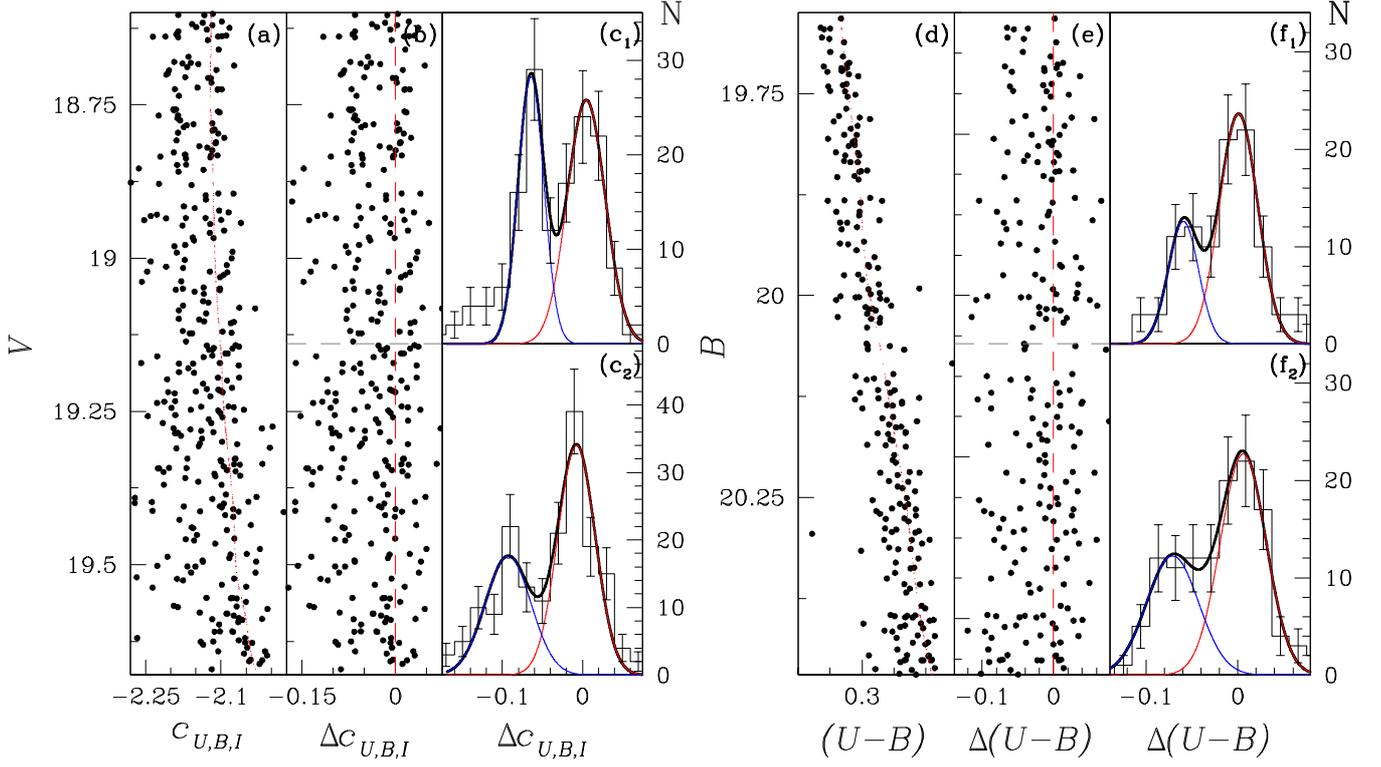}
  \caption{As in Fig.~\ref{n6752}, but for NGC~6121. In the case
      of $V$ versus $c_{U,B,I}$ CMD, we obtained that rMS and bMS
      contain respectively $60\pm7$\,\% and $40\pm8$\,\% of MS stars
      in panel (c1), $61\pm6$\,\% and $39\pm7$\,\% of MS stars in
      panel (c2). In the case of $B$ versus $(U-B)$ CMD we obtained
      that rMS a bMS contains respectively $67\pm7$\,\% and
      $33\pm10$\,\% of MS stars in panel (f1), $75\pm13$\,\% and
      $25\pm14$\,\% of MS stars in panel (f2).}
  \label{n6121}
\end{figure*}

We have already demonstrated that the $\Delta(U-B)$ distribution, in
the $B$ versus $U-B$ CMD of NGC\,6121, shows a double peak, proving
the presence of multiple populations (see panel a$_1$ of
Fig.~\ref{prove}). We performed a detailed analysis of the MS of
NGC\,6121, applying the same procedure described for NGC\,6752. The
procedure and the results are shown in Fig.~\ref{n6121}. We find from
the analysis of the $B$ versus $U-B$ CMD that the bMS contains
$31\%\pm8\%$ of MS stars, and the rMS includes the remaining
$69\%\pm7\%$. In the case of the $V$ versus $c_{U,B,I}$ diagram we
infer that rMS and bMS contain $61\%\pm4\%$ and $39\%\pm5\%$ of the
total number of MS stars, respectively. We computed the fraction of
rMS and bMS for NGC6121 using different histogram binsizes and
changing starting/ending points. We used binsizes with values between
0.005 and 0.025 (larger than color error), starting points between
-0.5 and -0.15 and ending points between 0.15 and 0.5. I all cases,
the resulting fraction of stars are in agreement with the values we
quote above within the errors The results from the two CMDs imply that
in the rMS and bMS there are, respectively, $63\%\pm3\%$ and
$37\%\pm4\%$ of MS stars.

\section{The radial distribution of stellar populations in NGC\,6752 and NGC\,6121\label{rad}}

\begin{figure*}
  \centering
  \includegraphics[bb=51 176 330 500, width=0.45\textwidth]{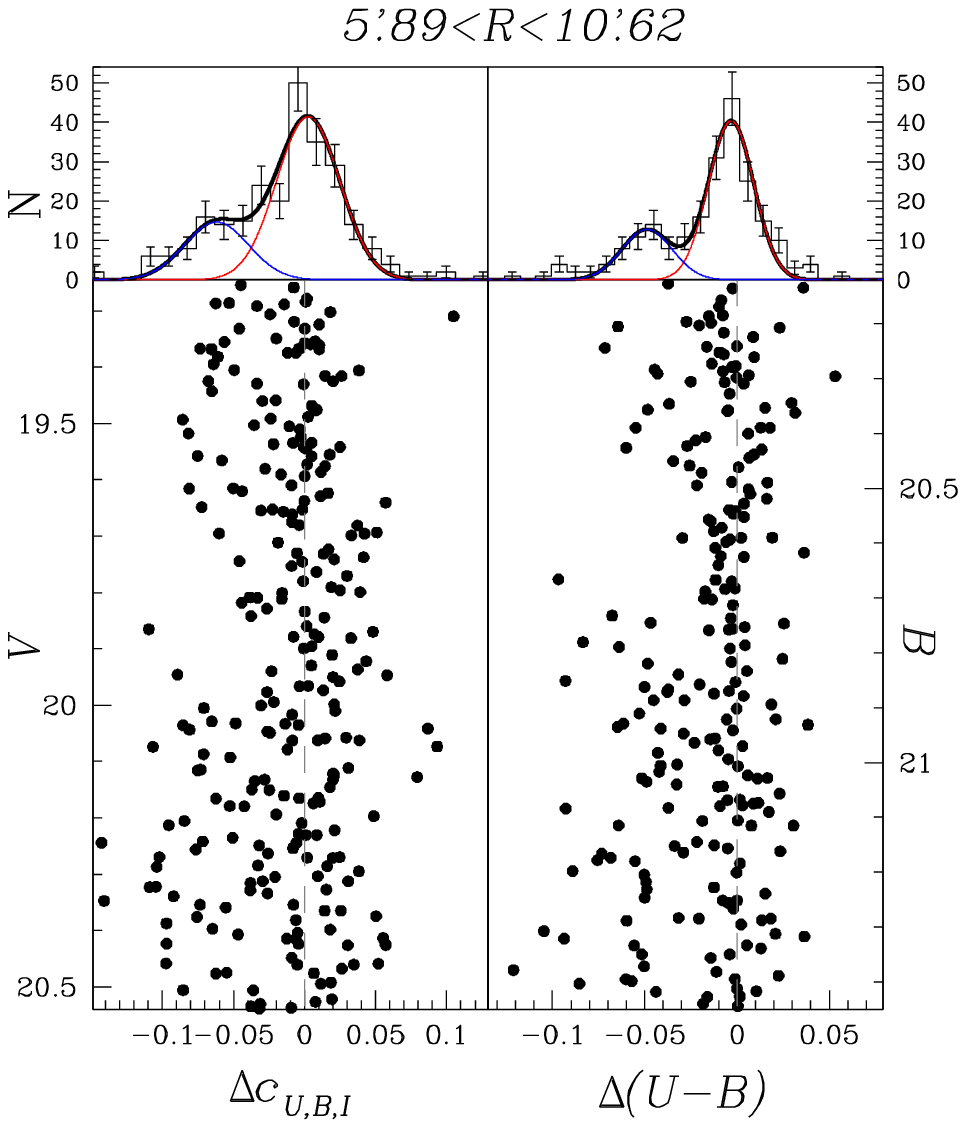}
  \includegraphics[bb=51 176 330 500, width=0.45\textwidth]{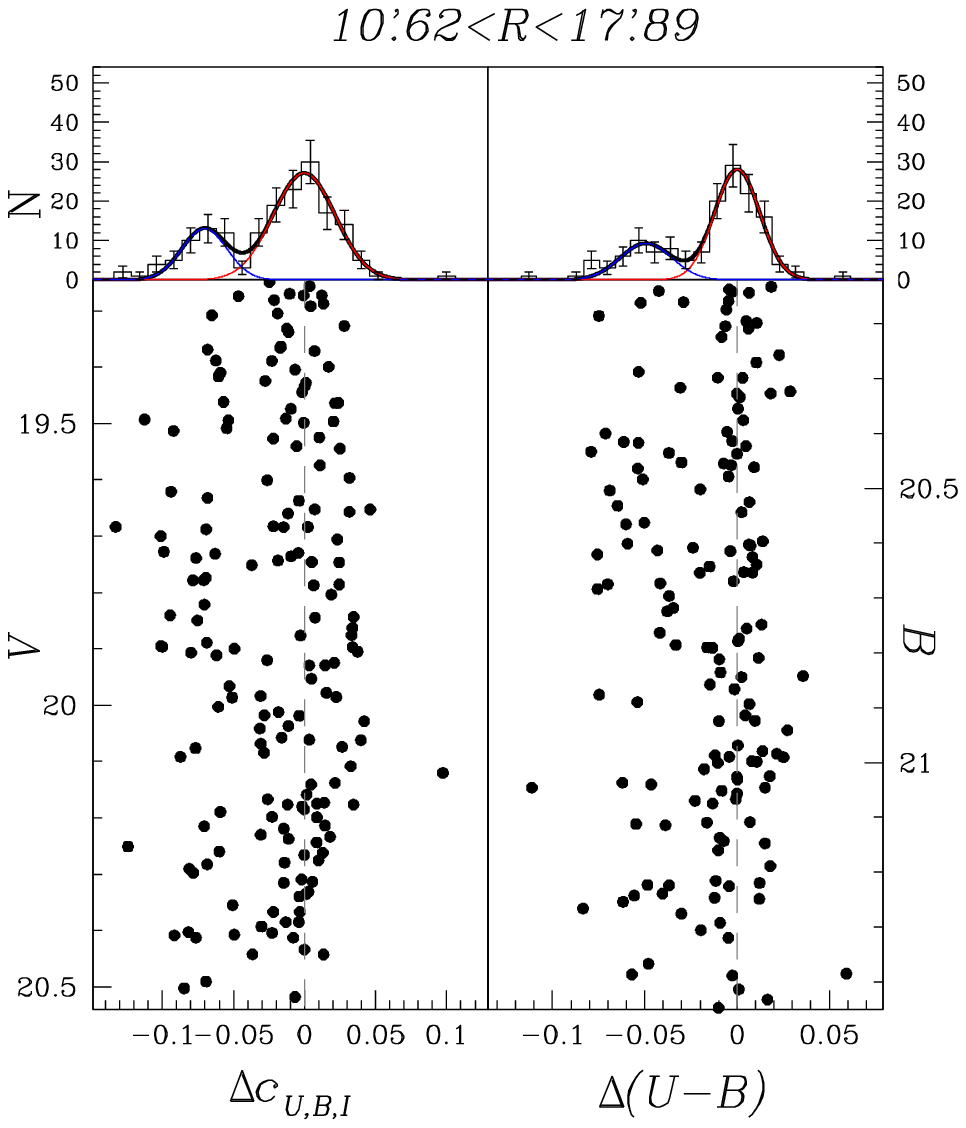}
  \caption{Color distribution analysis for the MSs stars of NGC~6752
    in two different radial bin, containing almost the same number
    (855 and 854) of stars.  In the inner field (left panels) we
      obtained that MSa (blue) and MSbc (red) contain respectively
      $28\pm6$\,\% and $72\pm4$\,\% of MS stars in the $V$ versus
      $c_{U,B,I}$ CMD, and $27\pm6$\,\% and $73\pm4$\,\% of MS stars
      in the $B$ versus $(U-B)$ CMD. In the outer field (right panels)
      we obtained that MSa and MSbc contain respectively $26\pm7$\,\%
      and $74\pm4$\,\% of MS stars in the $V$ versus $c_{U,B,I}$ CMD,
      and $29\pm8$\,\% and $71\pm6$\,\% of MS stars in the $B$ versus
      $(U-B)$ CMD.}
  \label{radn6752}
\end{figure*}

\begin{figure*}
  \centering
  \includegraphics[bb=74 290 340 610, width=0.45\hsize]{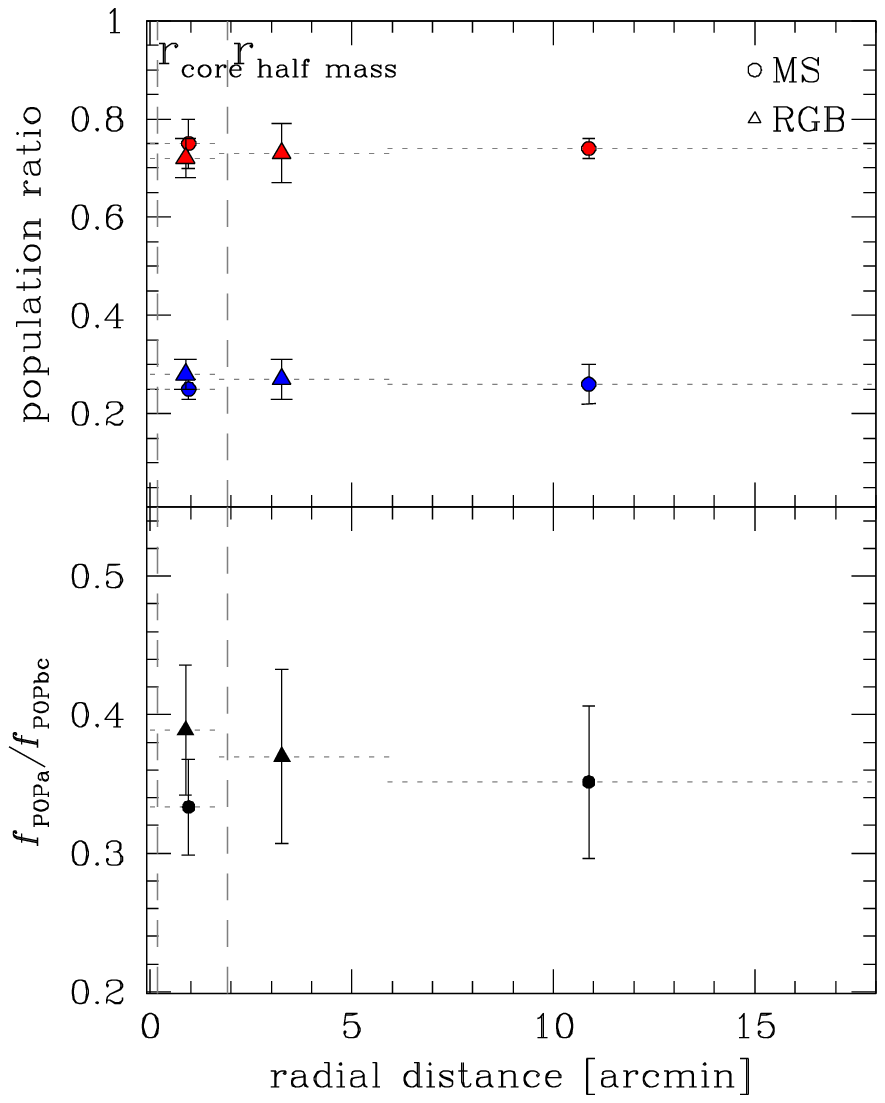} 
  \includegraphics[bb=74 290 340 610, width=0.45\hsize]{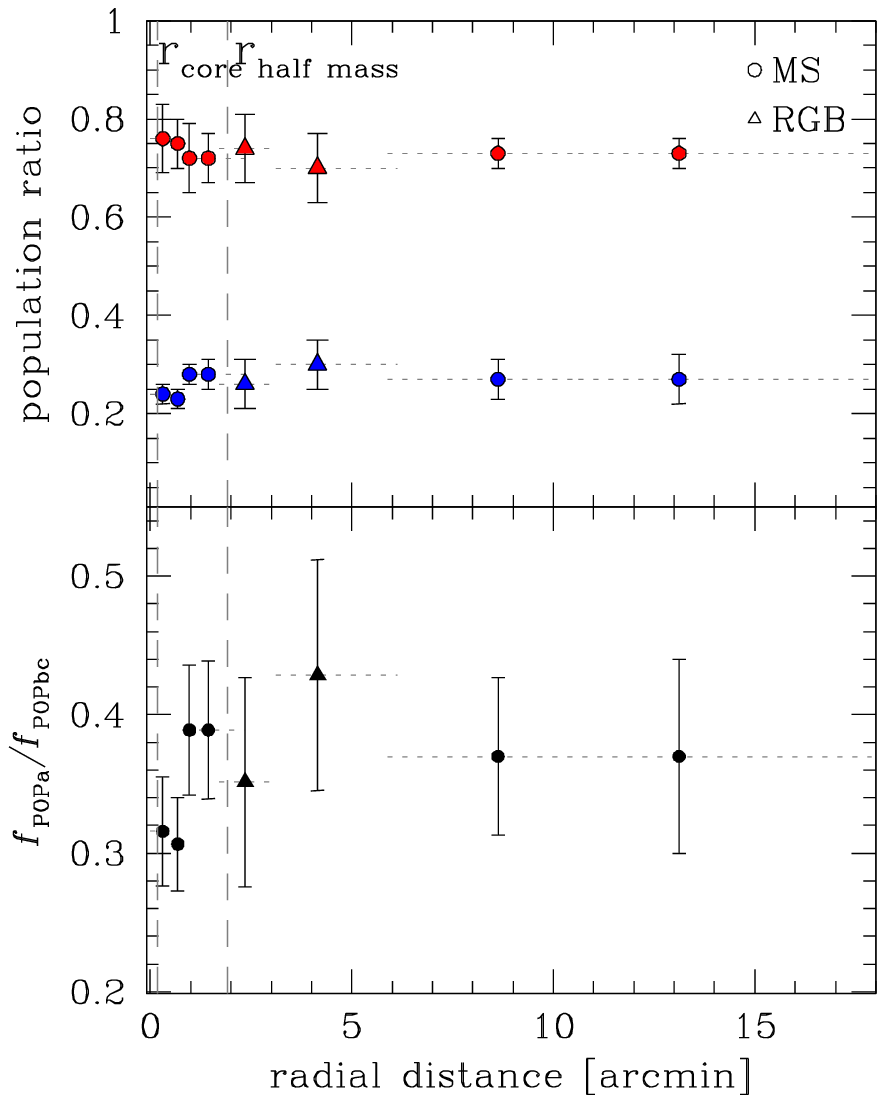}
  \caption{\textit{Top}: Radial distribution of the fraction of
    population `a' (blue) and `b'+`c' (red) stars with respect the total
    number of stars. \textit{Bottom}: radial trend of the ratio
    between $f_{\rm POPa}$ and $f_{\rm POPbc}$ stars. On the left we
    considered single radial interval for each set of data, while in
    the right panel we divided the radial interval in different bins.
    The distribution seems to be flat in both the cases.}
  \label{ratio}
\end{figure*}

The analysis of the radial distribution of rMS and bMS stars in
NGC\,6121 and NGC\,6752 is an important ingredient
to shed light on the formation and the evolution of multiple stellar
populations in these GCs.  Indeed, theoretical models predict that,
when the GC forms, second-generation stars should be more centrally
concentrated than first-generation ones, and many GCs could still keep
memory of the primordial radial distribution of their stellar
populations (e.g.\,\citealt{2008MNRAS.391..825D},
\citealt{2011MNRAS.412.2241B}, \citealt{2013MNRAS.429.1913V}).

The radial distribution of stellar populations in NGC\,6752 is still
controversial. \citet{2011A&A...527L...9K} determined wide-field
multi-band photometry of NGC\,6752 and studied the distribution of its
stellar populations across the field of view.  They have concluded
that there is a strong difference in the radial distribution between
the populations of RGB stars that are bluer (bRGB) and redder (rRGB)
in $(U - B)$ color, and obtained similar findings from the study of
the SGB.  Specifically, at a radial distance close to the half-mass
radius ($r_{\rm h} = 1\farcm91$; \citealt{1996AJ....112.1487H}, 2010
edition) the fraction of rRGB stars abruptly decreases.  These results
are in disagreement with the conclusions by Mi13 who showed that the
three stellar populations identified in their paper share almost the
same radial distribution.  Kravtsov and collaborators analyzed stars
with a radial distance from the center of NGC\,6752 out to $\sim
9\farcm5$ , while the study by Mi13 is limited to the innermost $\sim
6$\,arcmin. In this paper we extend the analysis to larger
radii\footnote {We assume that stars in the fields of NGC6752 and
    NGC6121 are representative of stellar populations at the studied
    radial distance; we are not able to investigate any dependency on
    the angular position using the dataset presented in this work.}.
 
As already mentioned in Sect. \ref{CMDs}, we suggest that the bMS of
NGC\,6752 corresponds to the population `a' identified by Mi13, while
the most populous rMS hosts both population `b' and population `c'
stars of Mi13. For this reason, in this section, we rename the bMS in MSa
and the rMS in MSbc.

In order to investigate the radial distribution of stellar populations
within the field of view analyzed in this paper, we divided the
catalog of NGC~6752 stars into two groups at different radial distance
from the cluster center, each containing almost the same total number of
stars.

The inner sample of stars (inner field)
  lies between $5\farcm89$ and $10\farcm62$ from the cluster center.
  The outer group of stars (outer field) is between $10\farcm62$ and
  $17\farcm89$ from the center.  We estimated the fraction of stars in
  each group by following the same procedure described in
  Sect.~\ref{popratio}.

The results are illustrated in Fig.~\ref{radn6752}. In the left panels
we show the verticalized $V$ versus $\Delta c_{U,B,I}$ and the $B$
versus $\Delta (U-B)$ diagrams for stars in the inner field.
In this region the MSa contains $27\% \pm 4\%$ and the MSbc hosts the
remaining $73\% \pm 3\%$ of the total number of MS stars.  In the
outer field (right panels of Fig. \ref{radn6752}) the MSa and the MSbc
are made of the $27\% \pm 5\%$ and $73\% \pm 3\%$ of MS stars,
respectively. We conclude that there is no evidence for a gradient
within the field of view studied in this paper.

To further investigate the radial distribution of stellar populations
in NGC\,6752 we compare the results obtained in this paper for stars
with distance from the cluster center larger than $\sim 6$\,arcmin and
the fraction of stars that have been estimated by Mi13 in the internal
regions by using the same method.

Since the MSbc contains both populations `b' and population `c' stars,
we have added together the fractions of population `b' ($f_{\rm
  POPb}$) and population `c' stars ($f_{\rm POPc}$) listed by
\citet[see their Table~4]{2013ApJ...767..120M} and calculated the
fraction of stars in these two populations: $f_{\rm POPbc}=f_{\rm
  POPb}+f_{\rm POPc}$.  As aforementioned in
Sect.~\ref{CMDs}, we further compare the fractions of population `a'
stars by Mi13, with the fractions of MSa stars derived in this paper.
The values of $f_{\rm POPa}$ and $f_{\rm POPbc}$ are listed in
Tab.~\ref{table:2}.

\begin{figure*}
  \centering
  \includegraphics[bb=51 176 330 500, width=0.42\textwidth]{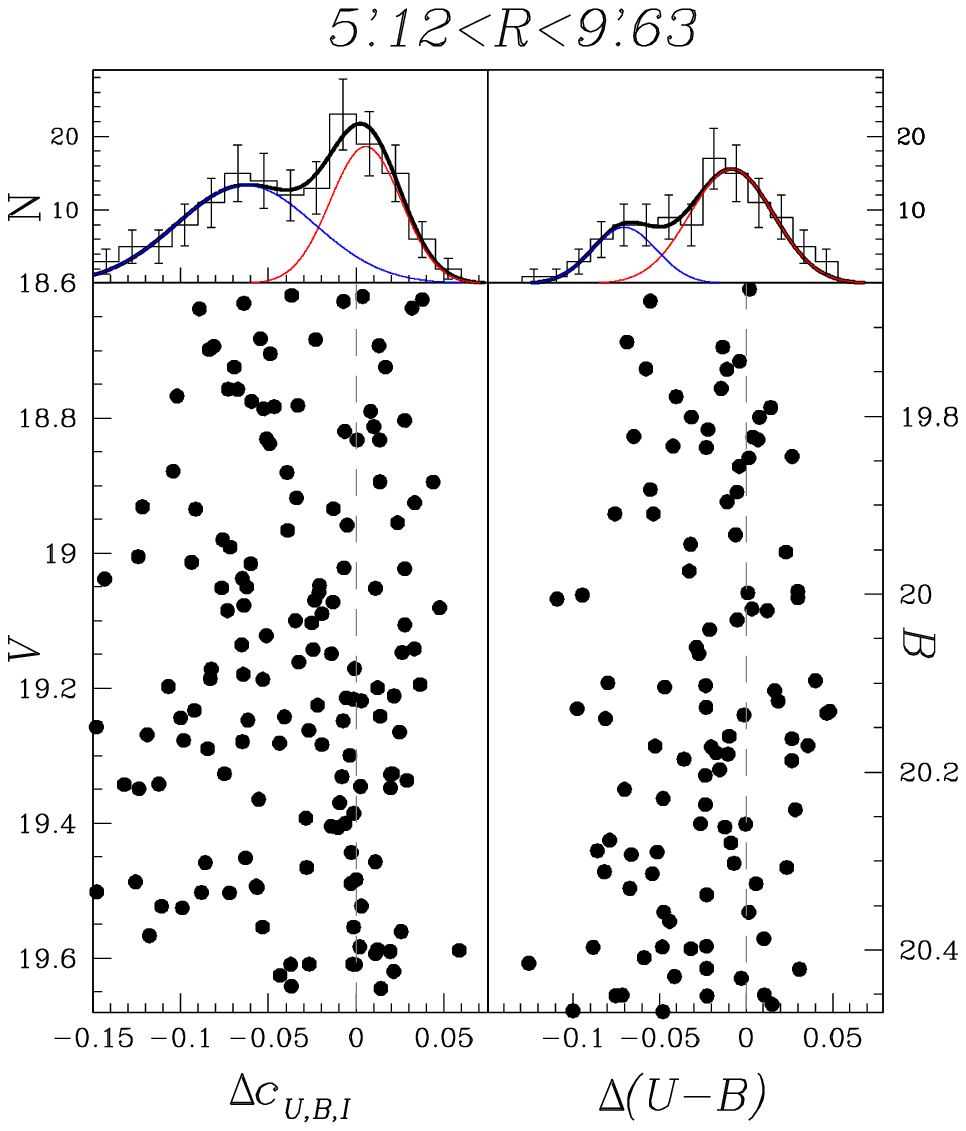}
  \includegraphics[bb=51 176 330 500, width=0.42\textwidth]{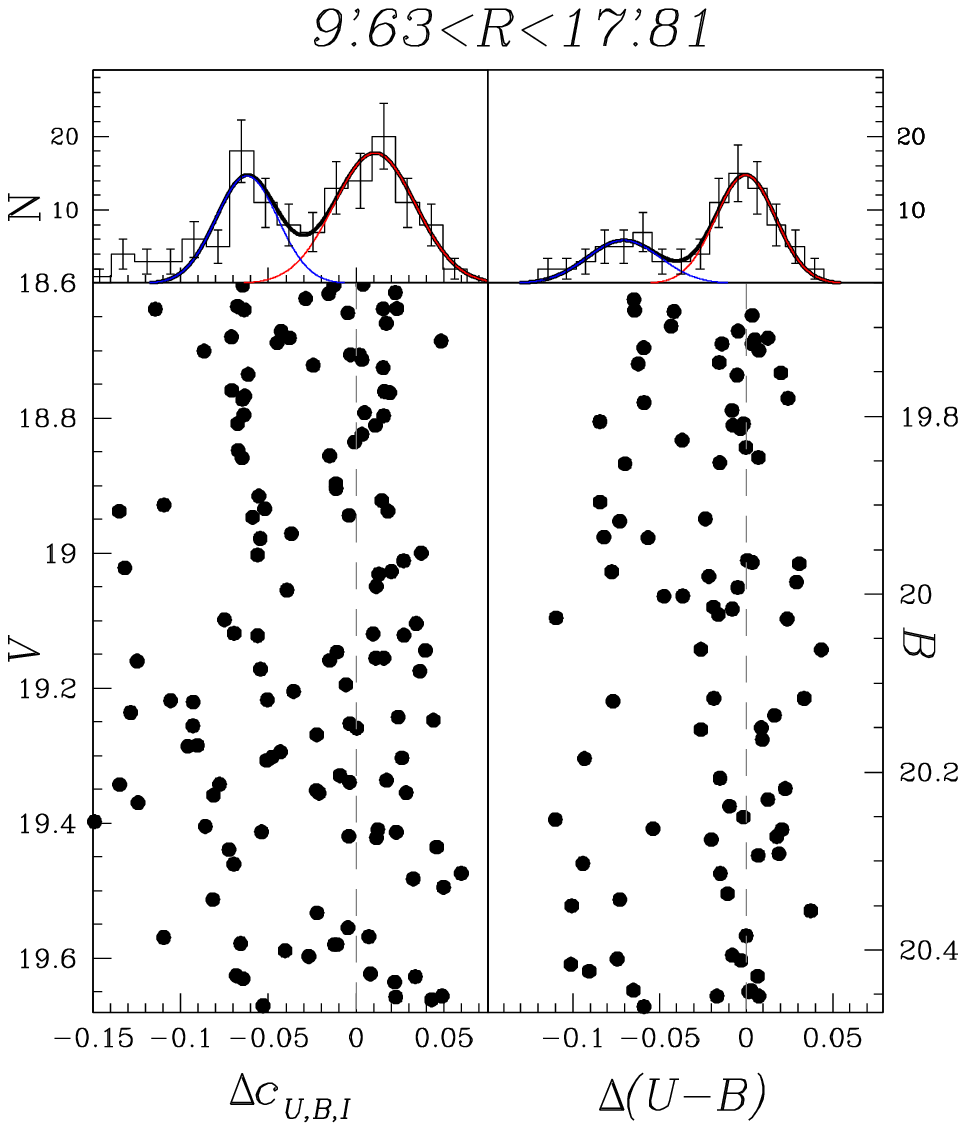}
  \caption{Color distribution analysis for the MSs stars of NGC~6121
    in two different radial bin, containing almost the same number
    (754 and 755) of stars.  In the inner field (left panels) we
      obtained that rMS and bMS contain respectively $42\pm20$\,\% and
      $58\pm20$\,\% of MS stars in the $V$ versus $c_{U,B,I}$ CMD, and
      $73\pm16$\,\% and $27\pm17$\,\% of MS stars in the $B$ versus
      $(U-B)$ CMD. In the outer field (right panels) we obtained that
      rMS and bMS contain respectively $62\pm6$\,\% and $38\pm7$\,\%
      of MS stars in the $V$ versus $c_{U,B,I}$ CMD, and $67\pm6$\,\%
      and $33\pm9$\,\% of MS stars in the $B$ versus $(U-B)$ CMD.}
  \label{radn6121}
\end{figure*}

\begin{table}
  \caption{Fraction of POPa and POPbc Stars for NGC\,6752}
  \label{table:2}  
  \centering       
  \begin{tabular}{c c c l l c}
    \hline\hline              
    $R_{\mathrm{min}}$ & $R_{\mathrm{max}}$ & $R_{\mathrm{ave}}$ & $f_{\rm POPa}$ & $f_{\rm POPbc}$ & Seq. \\
    \hline
    0.00 & 1.70 & 0.95 & $0.25 \pm 0.02 $ & $0.75 \pm 0.05$  & MS  \\
    0.00 & 1.70 & 0.87 & $0.28 \pm 0.03 $ & $0.72 \pm 0.04$  & RGB \\
    1.70 & 6.13 & 3.26 & $0.27 \pm 0.04 $ & $0.73 \pm 0.06$  & RGB \\
    5.89 &17.89 &10.88 & $0.26 \pm 0.04 $ & $0.74 \pm 0.02$  & MS  \\ 
    \hline
    0.00 & 0.53 & 0.31 & $0.24 \pm 0.02 $ & $0.76 \pm 0.07$  & MS  \\
    0.53 & 0.83 & 0.68 & $0.23 \pm 0.02 $ & $0.75 \pm 0.05$  & MS  \\
    0.83 & 1.12 & 0.97 & $0.28 \pm 0.02 $ & $0.72 \pm 0.07$  & MS  \\
    1.12 & 2.33 & 1.44 & $0.28 \pm 0.03 $ & $0.72 \pm 0.05$  & MS  \\
    1.70 & 3.11 & 2.35 & $0.26 \pm 0.05 $ & $0.74 \pm 0.07$  & RGB \\
    3.11 & 6.13 & 4.15 & $0.30 \pm 0.05 $ & $0.70 \pm 0.07$  & RGB \\
    5.89 &10.62 & 8.63 & $0.27 \pm 0.04 $ & $0.73 \pm 0.03$  & MS  \\ 
    10.62&17.89 &13.12 & $0.27 \pm 0.05 $ & $0.73 \pm 0.03$  & MS  \\ 
    \hline

  \end{tabular}
\end{table}

Results are shown in Fig~\ref{ratio}, where the top panels show the
distribution of the fraction of population `a' (in blue) and the
fraction of population `b'+`c' (in red) as a function of the radial
distance from the cluster center, while the bottom panels show the
radial trend of the ratio between the fraction of population `a' and the
fraction of population `b'+`c'. In the left panels we show both the above
described distributions considering single radial intervals for each
set of data, while in the right panels we divided each radial interval
in different bins.  Our findings suggest that there is no evidence for
a radial gradient among population `a' and population `b'+`c' of NGC\,6752.

In order to investigate the radial distribution of stellar populations
in NGC\,6121, we divided the field of view analyzed in this paper into
two regions, with radial distance from the cluster center
$5\farcm12<R<9\farcm63$ (inner field) and $9\farcm63<R<17\farcm81$
(outer field). Each region contains almost the same number of stars.
We determined the fraction of rMS and bMS stars by following the same
recipe described in detail for NGC\,6752. The results are shown in
Fig. \ref{radn6121}. We found that in the inner field the fraction of
bMS is $40\% \pm 13\%$ and the fraction of rMS is $60 \% \pm
13\%$. For the outer field we obtain that the bMS and the rMS contains
respectively the $36\% \pm 6\%$ and the $64\% \pm 4\%$ of the total
number of the considered MS. Also for NGC\,6121 we  found no evidence of population gradients.

\section{The helium content of stellar populations in  NGC\,6121 and NGC\,6752.}\label{Helium}

The ultraviolet pass-band is very efficient to separate multiple
sequences due to its sensitivity to difference in C, N, O abundance
(\citealt{2008A&A...490..625M}, \citealt{2011A&A...534A...9S}). In
contrast, $B-I$ and $V-I$ colors are marginally affected by
light-element variations, but are very sensitive to the helium
abundance of the stellar populations
(e.g.\,\citealt{2002A&A...395...69D}, \citealt{2007ApJ...661L..53P},
\citealt{2011A&A...534A...9S}, \citealt{2012AJ....144....5K},
\citealt{2013MmSAI..84...91C}), thus providing us with an efficient
tool to infer the helium content.
  
\subsection{NGC\,6121}
\begin{figure}
  \centering \includegraphics[bb=19 144 590 717,
    width=\hsize]{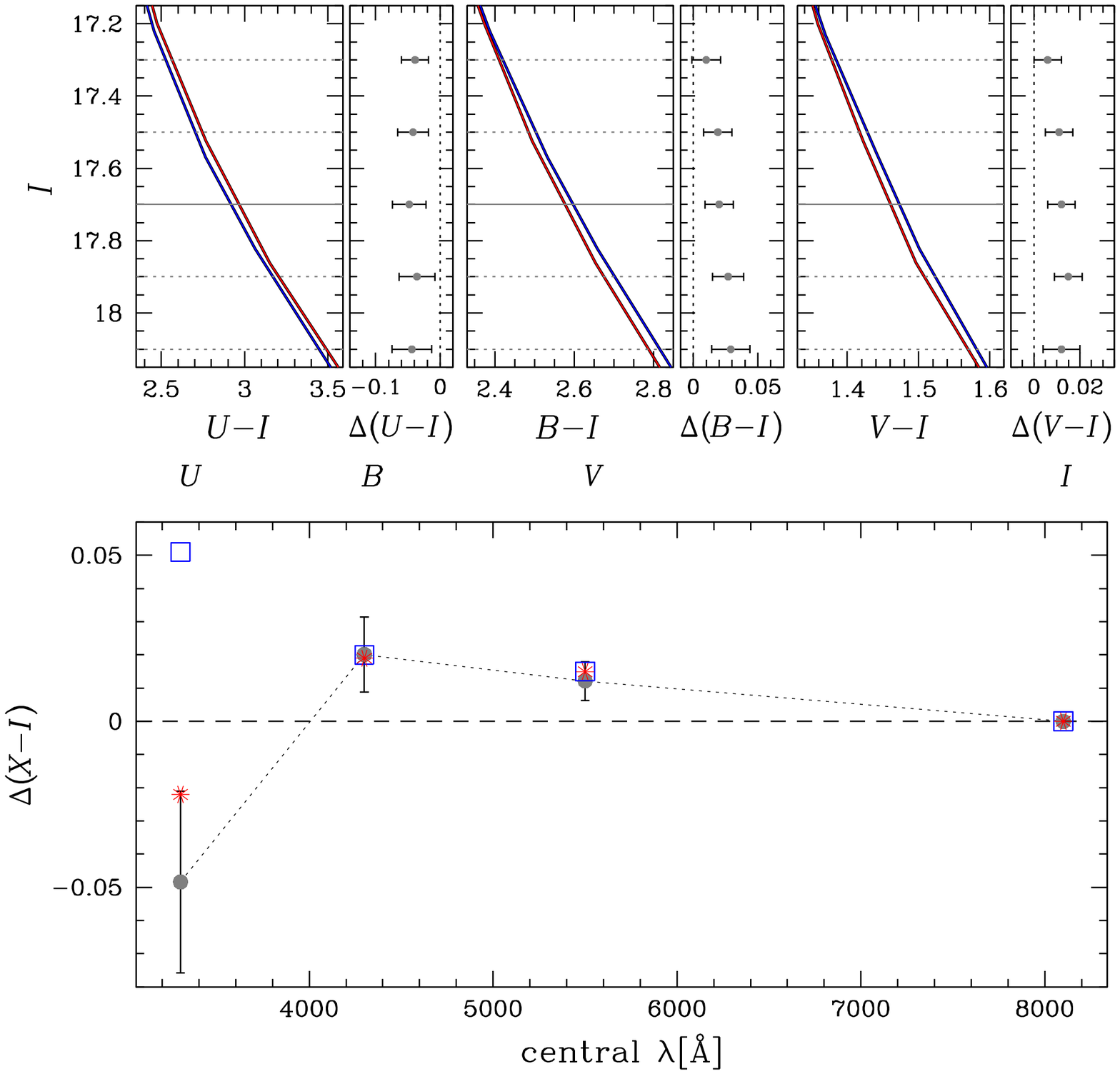}
  \caption{{\it Top panels:} MS fiducial in 3 $X-I$ CMDs ($X=U,B,V$
    ) of NGC\,6121. The horizontal lines represent the magnitudes
    $I_{\rm CUT}$ for which the color distance between the two MSs
    is calculated. For each $I_{\rm CUT}$ the error in
      $\Delta(X-I)$ is shown {\it Bottom panel:} $X-I$ color
    distance between rMS and bMS at $I_{\rm CUT}=17.7$ as a
    function of the central wavelength of the $X$
    filter. Observations are represented with gray dots. Red
    asterisks are the best-fitting model, while blue squares are
    the results obtained calculating synthetic colors of two MS
    stars with the same light-element chemical composition, but
    different He content. The blue squares demonstrate that the
    abundance of light elements assumed in the model does not
    affect the results on the He abundance of the two MSs in the
    optical colors but strongly affect the $U$ band. }

  \label{fig:elio6121}
\end{figure}

The procedure to estimate the average helium difference between bMS
and rMS stars for NGC\,6121 is illustrated in
Fig.~\ref{fig:elio6121}  and is already used
in several papers by our group (Mi13, \citealt{2012ApJ...745...27M,
  2012ApJ...744...58M}). Since we have already extracted the stellar
populations in NGC\,6121 by using the $B$ versus $U-B$ CMD of
Fig.~\ref{n6121}, we can now follow them in any other CMD.  By
combining photometry in four filters, we can construct three CMDs with
$I$ versus ($X-I$), where $X=U,B,V$. The fiducial lines of bMS and rMS
in these CMDs are plotted in the upper panels of
Fig.~\ref{fig:elio6121}. rMS is redder than bMS in $U-I$ color,
whereas it is bluer in $B-I$ and $V-I$ colors.

We measured the $X-I$ color distance between the two MSs at a
reference magnitude ($I_{\rm CUT}$), and repeated this procedure for
$I_{\rm CUT}$=17.3, 17.5, 17.7, 17.9, and 18.1 (corresponding to
  the magnitude interval where the two MS separations is maximal, cf
  Fig.~\ref{prove}). The color difference $\Delta$($X-I$) is plotted
in the lower panel of Fig.~\ref{fig:elio6121} as a function of the
central wavelength of the $X$ filter (gray dots), for the case of
$I_{\rm CUT}$=17.7.
 
We estimated effective temperatures ($T_{\rm eff}$) and gravities ($\log
g$) at different $I$=$I_{\rm CUT}$ for the two MS stars and for the
different helium contents by using
BaSTI isochrones (\citealt{2004ApJ...612..168P,2009ApJ...697..275P}).

We assumed  a
  primordial helium abundance for the bMS, $Y=0.248$, and used for
the rMS different helium content, with $Y$ varying from 0.248 to 0.400
in steps of $\Delta Y$=0.001.  To account for the appropriate chemical
composition of the two stellar populations of NGC\,6121 we assumed for
the bMS and the rMS the abundances of C, N, O, Mg, Al, and Na as
measured for first and second-generation RGB stars listed by
\citet[see their Table~6]{2008A&A...490..625M}.

We used the ATLAS12 program and the SYNTHE code
(\citealt{2005MSAIS...8...14K},\citealt{2005MSAIS...8...25C},
\citealt{2007IAUS..239...71S}) to generate synthetic spectra for the
adopted chemical compositions, from $\sim$2,500 \AA\ to $\sim$10,000
\AA. Synthetic spectra have been integrated over the transmission
curves of the $U,B,V,I$ filters, and,  we calculated the color difference ${\it X-I}$ for each
  value of helium of our grid.

The best-fitting model is determined by means of chi-square
minimization. Since the $U$ magnitude is strongly affected by the
abundance of light elements we used $B-I$ and $V-I$ colors only to
estimate $Y$.  The helium difference corresponding to the best-fit
models are listed in Table~\ref{tab:6121} for each value of $I_{\rm
  CUT}$.

We derived that the rMS is slightly helium enhanced with respect to
the bMS (which has $Y=0.248$), with an average helium abundance of
$Y=0.268\pm0.008$ This is the internal error estimated as the rms
scatter of the $N=5$ independent measurements divided by the square
root of $N-1$.  Results are shown in Fig.~\ref{fig:elio6121} for the
case of $I_{\rm CUT}$=17.7, where we represented the synthetic colors
corresponding to the best-fitting model as red asterisks.

For completeness we also calculated synthetic colors of two MS stars
with the $I=I_{\rm CUT}$ and the same chemical composition (same
abundance of light elements). We assumed for bMS primordial helium and
for rMS the helium abundance of the best-fitting model. Results are
represented as blue squares in Fig.~\ref{fig:elio6121} and confirm
that the abundance of light elements assumed in the model does not
affect our conclusion on the helium abundance of the two MSs, which
are based on the optical colors.  Instead the different CNO content
strongly affect the $U$ band.


In principle, the He content of stellar populations in GCs can
  also be estimated using He lines in HB star spectra (e.g. the HeI
  line at $\lambda=5875.6\AA$ line,
  \citealt{2009A&A...499..755V,2012ApJ...748...62V,2014MNRAS.437.1609M}
  ).  However, spectroscopic measurement of He in GC stars has many
  limitations.  First of all, He can only be measured for stars in a
  very limited temperature interval ($8500<T<11500$\,K).  In fact,
  stars with $T\le8500$\,K are not sufficiently hot to form He lines,
  while stars bluer than the Grundahl jump (\citealt{1999ApJ...524..242G},
  $T\ge11.500$K) are affected by He sedimentation and metal
  levitation which alter the original surface abundance.  The HB of
  NGC\,6121 is populated both on the red and the blue side of the RR
  Lyrae instability strip.  Spectroscopic investigation by
  \citet{2011ApJ...730L..16M} reveals that the blue HB is made of
  second population Na-rich and O-poor stars, while red HB stars
  belong to the first population.  In NGC\,6121, the HB segment with
  $8500<T<11500$ K corresponds to the blue HB, and therefore it only
  provides partial information only. In this cluster (as in many
  others) it is not possible to spectroscopically measure the He
  content of the first population.

  \citet{2012ApJ...748...62V} have used the HeI line at
  $\lambda=5875.6$ to estimate the helium content of six blue-HB stars
  in the blue HB of NGC\,6121. All of them are second-population
  stars.  They derived a mean value of $Y$=0.29$\pm$0.01 (random) $\pm$
  0.01 (systematic) and conclude that second-population stars would be
  enhanced in helium by $\sim$0.04-0.05 dex.  This estimate of the He
  content has been made by assuming LTE approximation.  However, the
  HeI line at $\lambda=5875.6$ is affected by NLTE effect, which can
  cause an error in the $Y$ estimate as large as $\Delta$\,$Y$=0.10 (see
  \citealt{2014MNRAS.437.1609M} for the case of NGC\,2808).  Appropriate
  NLTE analysis is required to infer reliable He abundances from
  spectroscopy of HB stars in NGC\,6121.  In contrast, the He
  difference between red- and blue-MS stars in NGC\,6121 comes from
  the colors of the fiducial lines, which have small color
  uncertainties.

\begin{table}
  \tiny
  \caption{Parameters used to simulate Synthetic Spectra of rMS and bMS stars and estimation of helium difference between the two population for different $I_{\rm CUT}$ in the case of NGC\,6121 }
  \label{tab:6121}  
  \centering       
  \begin{tabular}{cccccc}
    \hline\hline              
    $I_{\rm CUT}$ & $T_{\rm EFF, bMS}$ & $T_{\rm EFF, rMS}$ & $\log{g}_{\rm bMS}$ & $\log{g}_{\rm rMS}$ & $\Delta Y$ \\
    \hline
    17.3   &      5542   &      5571   &      4.58    &     4.57   &     0.014  \\
    17.5   &      5397   &      5444   &      4.60    &     4.60   &     0.021  \\
    17.7   &      5247   &      5297   &      4.63    &     4.63   &     0.022  \\
    17.9   &      5095   &      5149   &      4.65    &     4.65   &     0.024  \\
    18.1   &      4944   &      4989   &      4.66    &     4.66   &     0.021  \\
    \hline
    average &           &            &            &            &     0.020, $\sigma=$0.008\\
    \hline
  \end{tabular}
\end{table}


\subsection{NGC\,6752}

\begin{figure}
  \centering
  \includegraphics[bb=19 144 590 717, width=\hsize]{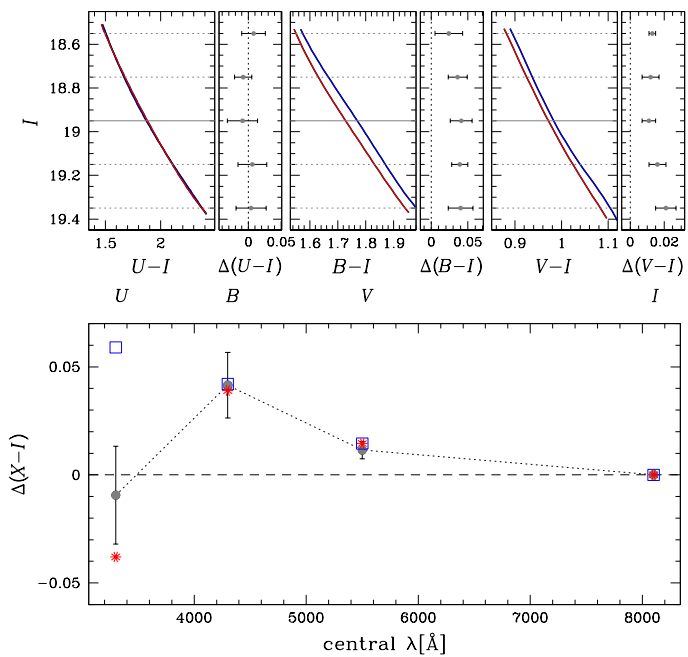}
  \caption{As in Fig.~\ref{fig:elio6121}, but for NGC\,6752. In the
    top panels, the fiducial blue is that one of the MSa and the
    red fiducial is for MSbc.}
  \label{fig:elio6752}
\end{figure}

We followed the same procedure to estimate the average helium
difference between MSa and MSbc stars. We measured the color distance
between the two fiducial lines of MSa and MSbc in the $I$ versus
$(X-I)$ CMDs (Fig.~\ref{fig:elio6752}), where $X=U,B,V$, at reference
magnitudes $I_{\rm CUT}=18.55,\,18.75,\,18.95,\,19.15$ and
$19.35$. The color difference $\Delta (X-I)$ at $I_{cut}=18.95$ is
plotted in the bottom panel of Fig.~\ref{fig:elio6752} as a function
of the central wavelength of the $X$ filter.

We used BaSTI isochrones
(\citealt{2004ApJ...612..168P,2009ApJ...697..275P}) to estimate
$T_{\rm eff}$ and $\log{g}$ at different $I_{\rm CUT}$.

We assumed that MSa has primordial helium abundance, $Y=0.248$, and
varied the helium content of the MSbc between 0.248 and 0.400 in steps
of $\Delta Y=0.001$. We assumed for the MSa the same C, N, O, Mg, Al
and Na abundances of the population `a' of Mi13; for the chemical
composition of the MSbc we considered the average of the abundances of
the population `b' and `c' listed by Mi13.

As mentioned above, we obtained synthetic spectra for the adopted
chemical compositions, integrated them over the transmission curves of
the $U$, $B$, $V$, $I$ filters, and computed the helium difference
using the best-fitting model.

We obtained that the MSbc is helium enhanced with respect to the MSa
of $\Delta Y=0.025 \pm 0.006$. As for NGC\,6121, since the $U$
magnitude is affected by the abundance of light elements, we used
$B-I$ and $V-I$ colors only to estimate $Y$. Note that the
abundance of light elements assumed in the model does not affect our
conclusion on the helium abundance of the two MSs, as already proved
in the case of NGC\,6121.

\begin{table}
  \tiny
  \caption{Parameters used to simulate Synthetic Spectra of MSa and MSbc stars and estimation of helium difference between the two population for different $I_{\rm CUT}$ in the case of NGC\,6752}
  \label{tab:6752}  
  \centering       
  \begin{tabular}{cccccc}
    \hline\hline              
    $I_{\rm CUT}$ & $T_{\rm EFF, MSa}$ & $T_{\rm EFF, MSbc}$ & $\log{g}_{\rm MSa}$ & $\log{g}_{\rm MSbc}$ & $\Delta Y$ \\
    \hline
        18.55   &     5410  &      5456  &      4.65  &      4.65  &     0.021  \\
        18.75   &     5254  &      5306  &      4.67  &      4.67  &     0.023  \\
        18.95   &     5100  &      5150  &      4.69  &      4.69  &     0.023  \\
        19.15   &     4946  &      4998  &      4.70  &      4.70  &     0.025  \\
        19.35   &     4798  &      4851  &      4.72  &      4.72  &     0.026  \\
\hline
       average &           &            &            &            &     0.024, $\sigma=$0.006  \\
    \hline
  \end{tabular}
\end{table}

\subsection{Relation between HB morphology and Helium abundance}
In their work, \citet{2014ApJ...785...21M} have sought correlations
between HB morphology indicators and physical and morphological GC
parameters. Among these parameters there is also the maximum helium
difference between stellar populations hosted by GCs.

They introduced two different parameters to describe the HB
morphology: $L_1$, that is the color difference between the RGB and
the coolest border the HB, and $L_2$, that is the color extension of
the HB (for more details, see Fig.~1 of \citealt{2014ApJ...785...21M}).

They divided the sample of 74 GCs in three groups: in the first group,
G1, there are GCs with [Fe/H]$\ge -1.0$; the second group, G2,
includes GCs with [Fe/H]$< -1.0$ and $L_1\le 0.4$; the third group, G3 contains
GCs with $L_1> 0.4$.

They found a tight correlation between $L_2$ and the maximum
internal helium difference ($\Delta Y$, measured on the MS) for the group
G2+G3 (see Fig.~8 of their paper).

In our work we add two more points to their data-set, the Helium
difference between the two populations of NGC\,6752 and NGC\,6121, as
computed in this work.
In the case of NGC\,6752, the added point constitutes a lower limit
because the Helium difference between Pop$_{\rm a}$ and Pop$_{\rm
  bc}$, $\Delta Y$(Pop$_{\rm a}$-Pop$_{\rm bc}$), is the average value
between $\Delta Y$(Pop$_{\rm a}$-Pop$_{\rm b}$) and $\Delta
Y$(Pop$_{\rm a}$-Pop$_{\rm c}$).

\begin{figure}
  \centering
  \includegraphics[bb=180 290 370 500, width=0.7\hsize]{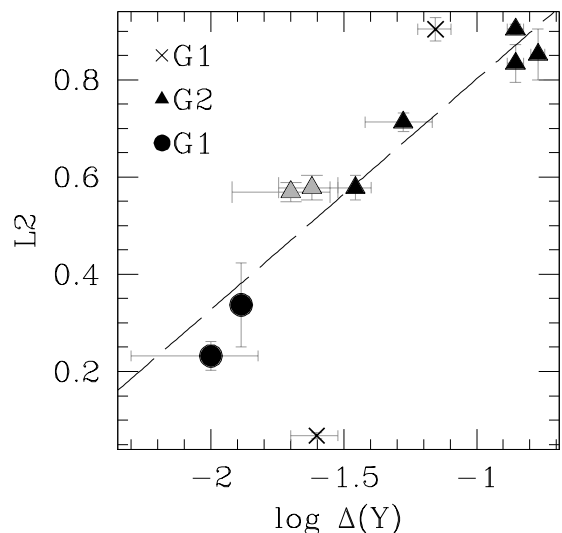}
  \caption{The HB morphological parameter $L_2$ as a function of the
    logarithm of the maximum helium difference among stellar
    populations in GCs. The black line is the best-fitting straight
    line for G2+G3 GCs. In black there are the data of
    \citet{2014ApJ...785...21M}, in grey the data of this work.}
  \label{hb}
\end{figure}

The result is in Fig.~\ref{hb}: in black there are the points of
\citet{2014ApJ...785...21M} and in grey the points added in this
work. In analogy to the work of \citet{2014ApJ...785...21M}, the
crosses refer to the G1 GCs, triangles to G2 group and dots to G3
clusters. Our data points confirm the tight correlation between
L2 and $\Delta Y$. We found a Spearman's rank correlation coefficient
$r_{\rm G2+G3}=0.93$ (to be compared to $r_{\rm G2+G3}=0.89 \pm0.17$
found by \citealt{2014ApJ...785...21M}), with $\sigma_{r,{\rm
    G2+G3}}=0.08$ (the uncertainty in $r$ is estimated by means of
bootstrapping statistic, as in \citealt{2014ApJ...785...21M}).

This result is a further proof that the helium-enhanced stellar
populations are likely related to the HB extension,
as predicted by theory.

\section{Summary} \label{sum}

The photometric analysis of ESO/FORS2 data of the external regions of
the three nearby Galactic GCs NGC\,6121 (M\,4), NGC\,6752 and
NGC\,6397 has confirmed that the first two GCs
host multiple stellar populations. Indeed, the $B$ versus $U-B$ and
$V$ versus $c_{U,B,I}$ CMDs of NGC\,6752 and NGC\,6121 show a split of
the MS in two components. Excluding the unique case of $\omega$\,Cen,
this is the first time that a split of the MS is observed using
ground-based facilities.

The multiple stellar populations of NGC\,6397 was investigated by
\citet{2012ApJ...745...27M} using {\it HST} data. They found two
stellar populations characterized by a modest helium variation $\Delta
Y\sim 0.01$. Unfortunately, in this work, it was not possible to
analyze these populations, because of the size of our photometric errors is
  comparable to the small color separation between the MSs.

Using {\it HST} data, Mi13 have already demonstrated that NGC\,6752
host three stellar populations. They computed the radial trend of the
ratio between the number of stars of different populations out a
radial distance from the center of $6\farcm13$. Because of larger
photometric errors, we have resolved only two MSs. Comparing them with
the work of Mi13, we found that the less populous MS corresponds to
their population `a', while the most populous MS hosts both their
populations `b' and `c'. In average we found that the MSa contains
about 26\% of the total number of stars and the MSbc host about 74\%
of the MS stars. The most straightforward interpretation is that the
MSa is formed by stars of the first generation with chemical
abundances similar to that of the Galactic halo field stars with the
same metallicity; the MSbc hosts stars of second generation, formed
out of material processed through first-generations stars. This
population is characterized by stars enhanced in helium, with $\Delta
Y = 0.025$. Our measurement of the helium enhancement is in agreement
with the average $\Delta Y$ of the populations `b' ($\Delta Y \sim
0.01$) and `c' ($\Delta Y \sim 0.03$) obtained by Mi13. We extended
the study of the radial trend of the populations of NGC\,6752 to more
external regions, confirming the results of Mi13, of a flat
distribution. Therefore we cannot confirm the results by
\citet{2011A&A...527L...9K} and \citet{2014ApJ...783...56K}; they
found that the two populations show a strong gradient at a radial
distance close to the half-mass radius.

In a recent work on NGC\,6121, \citet{2014MNRAS.439.1588M} investigate
the bottom of the MS of this cluster using {\it HST} near-infrared
photometry. They found that the MS splits into two sequences below the
MS knee. In particular they identified two MSs: a MS$_{\rm I}$ that
contains $\sim 38\%$ of stars and MS$_{\rm II}$ formed by the
remaining $62\%$. They show that the split of the MS is mainly due to
the effect of ${\rm H_2O}$ molecules, present in the atmospheres of
M-dwarfs, on their near-infrared color, and that it is possible to
associate the MS$_{\rm I}$ to a first generation of stars and the
MS$_{\rm II}$ to a second one. \citet{2008A&A...490..625M}, analyzing
spectra of RGB stars, found that $\sim 64\%$ of stars are Na-rich and
O-poor and the remaining $\sim 36\%$ have chemical abundances similar
to those of Halo-field stars with the same metallicity. All these
results are in agreement with what we have obtained in this work: the
MS of NGC\,6121 splits in the $B$ versus $(U-B)$ and $V$ versus
$c_{U,B,I}$ CMDs. We found two MSs: a less populous MS that contains
$\sim 37\%$ of MS stars and which constitutes the first generation of
stars and a more populous second generation MS that contains $\sim
63\%$ of stars. \citet{2012ApJ...748...62V}, using spectroscopic
measurements of blue HB (bHB) stars, obtained that the difference in
helium abundance between these stars and the red HB (rHB) stars is
$\Delta Y=0.02/0.03$. A spectroscopic analysis of
\citet{2011ApJ...730L..16M} revealed that the rHB stars have
solar-scaled [Na/Fe], while bHB stars are Na enhanced. In contrast to
the results of \citet{2012ApJ...748...62V}, a lower constraint to the
level of He enhancement is set by \citet{2014ApJ...782...85V},
founding a maximum $\Delta Y=0.01$ between bHB and rHB
stars. Analyzing how the two MS of NGC\,6121 behave in different CMDs,
we computed the helium abundance difference between them. Our result
is $\Delta Y = 0.020 \pm 0.005$, in agreement with that obtained by
\citet{2012ApJ...748...62V}. Also in the case of NGC\,6121, we did not
find evidence of changes in the fraction of bMS/rMS stars in the
radial range between $1.2\,r_{\rm h}<R<4.1\,r_{\rm h}$.

\citet{2014ApJ...785...21M} found a correlation between the HB
morphological parameter $L_2$ and the maximum helium difference among
stellar populations in GCs. Using the helium abundances computed in
this work for NGC\,6121 and NGC\,6752, we confirm this correlation and
the theoretical indications that helium enhanced stellar populations
are responsible of the HB extension.

\begin{acknowledgements}
DN is supported by a grant ``Borsa di studio per l’estero, bando
2013'' awarded by ``Fondazione Ing.  Aldo Gini'' in Padua (Italy).
APM acknowledges the financial support from the Australian Research
Council through Discovery Project grant DP120100475. AFM has been
supported by grants FL110100012 and DP120100991.
\end{acknowledgements}

   \bibliographystyle{aa} 
   \bibliography{biblio}


 \end{document}